\documentclass[11pt,a4paper]{iopart}
\usepackage[utf8]{inputenc}
\expandafter\let\csname equation*\endcsname\relax

\expandafter\let\csname endequation*\endcsname\relax 
\usepackage{amsmath}
\usepackage{amsfonts}
\usepackage{amssymb}
\usepackage{makeidx,bm}
\usepackage{graphicx}
\usepackage{iopams}
\usepackage{todonotes}
\usepackage{hyperref}

\usepackage{soul}

\def\bea{\begin{eqnarray}}
\def\eea{\end{eqnarray}}

\def\la{\langle}
\def\ra{\rangle}

\newcommand{\ab}{\textrm{AB}}
\newcommand{\rt}{\textrm{RT}}
\newcommand{\dr}{\textrm{DR}}
\newcommand{\ce}{\textrm{ce}}
\numberwithin{equation}{section} 

\usepackage{etoolbox}
\makeatletter
\newrobustcmd{\fixappendix}{%
  \patchcmd{\l@section}{1.5em}{7em}{}{}%
  \patchcmd{\l@subsection}{2.3em}{7em}{}{}%
}
\makeatother

\begin{document}

\title{Universal framework for the long-time position distribution of free active particles}

\author{Ion Santra$^{1}$, Urna Basu$^{1,2}$, Sanjib Sabhapandit$^{1}$}

\address{$^{1}$Raman Research Institute, Bengaluru 560080, India}

\address{$^{2}$S. N. Bose National Centre for Basic Sciences, Kolkata 700106, India}

\begin{abstract}
Active particles self-propel themselves with a stochastically evolving velocity, generating a persistent motion leading to a non-diffusive behavior of the position distribution. Nevertheless, an effective diffusive behavior emerges at times much larger than the persistence time. Here we develop a general framework for studying the long-time behaviour for a class of active particle dynamics and illustrate it using the examples of run-and-tumble particle, active Ornstein-Uhlenbeck particle, active Brownian particle, and direction reversing active Brownian particle. Treating the ratio of the persistence-time to the observation time as the small parameter, we show that the position distribution generically satisfies the diffusion equation at the leading order. We further show that the sub-leading contributions, at each order, satisfies an inhomogeneous diffusion equation, where the source term depends on the previous order solutions. We explicitly obtain a few sub-leading contributions to the Gaussian position distribution. As a part of our framework, we also prescribe a way to find the position moments recursively and compute the first few explicitly for each model. 
\end{abstract}

\noindent\rule{\hsize}{2pt}
\tableofcontents
\noindent\rule{\hsize}{2pt}
\title[Universal framework for the long-time position distribution of free active particles]
\maketitle

\section{Introduction}
There has been a growing interest in exploring various aspects of active particles in the recent years~\cite{Mori2021,Rosalba2021,Zhang2022,Santra2020_reset,Squarcini2022,Mori2020_prl,Hartmann2020,Singh2020,Maes2018,Woillez2019,Banerjee2020}. Active particles refer to self-propelled agents which can generate persistent motion by extracting energy from their surroundings at an individual level~\cite{MarchettiRev,BechingerRev,SriramRev}. Examples of such motion are found in living systems ranging from bacteria~\cite{bergbook} at the microscopic scale to the flocking of birds~\cite{Giardina2014,Bialek2012} and fish schools~\cite{Partridge1982,Guttal2020} at the macroscopic scale as well as in artificial systems including Janus particles~\cite{Sano2010,Bechinger2012}, and granular media~\cite{Tsimring2008,Sriram2014}. Constructing stochastic models plays a major role in the theoretical attempts to understand active motion. Run-and-tumble particle (RTP)~\cite{Berg1972,Tailleur2008,Malakar2017,Santra2020}, active Ornstein-Uhlenbeck particle (AOUP)~\cite{Leonardo2014,Martin2021}, active Brownian particle (ABP)~\cite{Howse2007,Basu2018}, and direction reversing active Brownian particle (DRABP)~\cite{Thutupalli2019,Santra2021} are among the most studied theoretical models, due to their simplicity as well as wide-ranging applicability in real physical systems. The common feature of these models is that they all describe the motion of an overdamped particle with a fluctuating  propulsion velocity that is correlated in time---effectively generating a persistent motion. They, however, differ in the stochastic dynamics of the velocity, modeling  different physical situations. The RTP, for example, models a typical bacterial motion moving with a constant speed, interrupted by intermittent tumbling events changing the direction of the velocity randomly. The velocity direction for an ABP, on the other hand, undergoes a rotational diffusion, mimicking the motion of some Janus particles. The DRABP models a certain type of bacterial motion, where the velocity undergoes a complete directional reversal intermittently, in addition to the ABP dynamics. Finally, each component of the velocity undergoes an Ornstein-Uhlenbeck process for AOUP, which due to its simplicity, has been used widely to investigate the nonequilibrium nature of active motion~\cite{Fodor2016}. 

A remarkable feature of active particles is that they show many intriguing behavior even at the single particle level. For example, in the presence of confining potentials, the position of an active particle typically has a non-Boltzmann-Gibbs distribution~\cite{Pototsky2012,Dhar2019,Malakar2020,Basu2019,Basu2020,Santra2021_trap}. Even in the absence of confining potential, striking signatures of activity show up at times shorter than the intrinsic persistence time-scales~\cite{Santra2020,Basu2018,Santra2021,Meerson2020}. At late-times, however, one expects an active particle to show diffusive behavior similar to a passive Brownian particle~\cite{MarchettiRev,BechingerRev}. This eventual diffusive behavior can be heuristically explained by expressing the total displacement $x(t)$ during a time interval $[0,t]$ as $x(t)=\sum_{i=1}^N\Delta x_i$, with $\Delta x_i$ being the increment over the time interval $[(i-1)\Delta t,i\Delta t]$, where $\Delta t=t/N$ is chosen to be much longer than the persistence time-scale. Consequently, ignoring the correlations among $\{\Delta x_i\}$, a diffusive Gaussian distribution for $x$ in the large $N$ limit can be anticipated by appealing to the central limit theorem. However, an exact systematic derivation of the diffusion equation corresponding to the Gaussian distribution, which is expected to be more involved, is still lacking. Moreover, we anticipate some signatures of activity to show up as sub-leading corrections to the Gaussian~\cite{Basu2018,Santra2020}, which cannot be obtained from the heuristic argument above.

  The goal of the manuscript is to develop a unified framework to describe the long-time behavior of the basic active particle models. To this end, we develop a systematic perturbative procedure to solve the master or the Fokker-Planck equation in the long-time regime, treating the ratio of the persistence-time to the observation time as the small parameter. We illustrate this for the four basic active particle models mentioned above. We find that the position distribution satisfies the diffusion equation at the leading order for all the models. We further show that the sub-leading contributions, at each order, generically satisfy an inhomogeneous diffusion equation, where the source term depends on the previous order solutions. We explicitly solve them for the first few orders and obtain the sub-leading contributions to the position distribution exactly. As a part of our framework, we also prescribe a way to find the position moments recursively and compute the first few explicitly for each model.

 Incidentally, the exact position distributions at all times are known in closed forms for RTP~\cite{Malakar2017} and AOUP (see~\ref{App:aoup}), as an infinite series in the Fourier space for the ABP~\cite{Basu2019,Franosch2016}. 
However, the commonalities between the large-time behavior of the different active particle models are not obvious from the known exact solutions and become apparent only through our analysis. The emphasis of this work is not on the explicit final results but rather the generic framework, which can be used to extract the long time behavior of active particle models, even when the exact solution is not known. In fact, we use the exact results of these three models as a test bed for validating our framework. Subsequently, we use the framework for DRABP to obtain the Gaussian distribution that was known only heuristically~\cite{Santra2020}, and the subleading corrections that were not known previously.

The procedure followed here falls within the general perturbative scheme of dealing with differential equations involving small parameters. For example, starting from the Kramers equation, the Fokker-Planck equation of a Brownian particle in the presence of a potential can be obtained in the overdamped limit (see Chapter VIII, Sec.~7 of~\cite{vanKampenbook}). Another well-known example is the derivation of the Fokker-Planck equation for slow degrees of freedom in an interacting many particle system, by integrating out the fast degrees~\cite{vanKampen1986,Bhat2019}. Recently, similar procedures were used in the context of active particles, for obtaining a perturbative expansion of the stationary states of AOUP~\cite{Martin2021} and ABP~\cite{Malakar2020} in external potentials about the respective passive Boltzmann distributions.

The paper is organized as follows. We first sketch the perturbative procedure in generic terms in Sec.~\ref{Sec:genstrategy} and outline our main results for the benefit of the readers. The explicit calculations for the position distributions, as well as the moments, are presented in Sec.~\ref{Sec:RTP}--\ref{Sec:DRABP} for the specific models RTP, AOUP, ABP and DRABP, respectively. Finally, we conclude with some general remarks in Sec.~\ref{conclusion}. We have moved some of computation details to the Appendices for better readability.

%%%%%%%%%%%%%%%%%%%
\section{Perturbative framework and main results}\label{Sec:genstrategy}

In this section, we briefly outline the main steps of the general perturbative framework for a one-dimensional active motion. The details, of course, depend on the specific model under consideration. The full picture will become clear in the subsequent sections where we explicitly carry out this perturbative procedure for the different models. 

The active particle models under consideration are generically described by the overdamped Langevin equation,
\bea
\dot{x}(t)=v(t),
\eea
where the propulsion velocity $v(t)$ independently evolves by a stochastic dynamics with a characteristic time $\tau_0$. In all the models considered in this paper, $v$ eventually reaches a stationary state with an exponentially decaying autocorrelation function $\la v(t)v(t')\ra\propto \exp(-|t-t'|/\tau_0)$. 

The joint distribution $P(x,v,t)$ satisfies a Fokker-Planck or a master equation,
\bea
\frac{\partial P}{\partial t}=-v\frac{\partial P}{\partial x}+\mathcal{L}_v P,
\eea
where the specific form of the operator $\mathcal{L}_v$ corresponding to the stochastic dynamics of $v$ depends on the specific model. We expand the joint distribution $P(x,v,t)$ as,
\bea
P(x,v,t)=\sum_{n=0}^\infty \psi_n(v)\,F_n(x,t),
\label{gen:basis}
\eea
where $\psi_n(v)$ are the eigenfunctions of $\mathcal{L}_v$ with $\psi_0(v)$ denoting the stationary state of $v$ satisfying $\mathcal{L}_v\psi_0(v)=0$. Evidently, the position distribution is given by,
\bea
\rho(x,t)=\int dv P(x,v,t)= F_0(x,t),
\eea
as $\int dv\, \psi_n(v)=\delta_{n,0}$. Note that, $\int dv$ can also indicate a sum over possible discrete states as in RTP. For the sake of simplicity, we choose our initial conditions such that the position distribution is even, i.e., $\rho(x,t)=\rho(-x,t)$, at all times.

We show that when the initial propulsion velocity $v(0)$ is chosen from the stationary state $\psi_0(v)$, the marginal position distribution admits the series expansion in the dimensionless small perturbation parameter $\tau_0/t$,
\bea
\rho(x,t)\equiv F_0(x,t)=\sum_{k=0}^\infty \tau_0^k\, A_0^{2k}(x,t),
\label{gen:strategy}
\eea
where $t^{-k}$ is absorbed in the series coefficient $ A_0^{2k}(x,t)$ for computational convenience [see \eref{gen:a02k_ansatz} below].
The choice of the superscript $2k$ in the notation is essentially related to the fact that $A_0^{2k}(x,t)$ is an even function of $x$. It will become clear when we show the explicit calculation in later sections.
 We find that the leading term $A_0^0(x,t)$ always satisfies the diffusion equation,
\bea
\frac{\partial A_0^0}{\partial t}=D_{\text{eff}}\frac{\partial^2 A_0^0}{\partial x^2},
\label{g:diffeq}
\eea
resulting in the familiar long-time Gaussian distribution,
\bea
A_0^0(x,t)=\frac{1}{\sqrt{4\pi D_{\text{eff}}t}}\exp\left(-\frac{x^2}{4D_{\text{eff}}t}\right).\label{g:diffinhomo}
\eea
The explicit form of the effective diffusion coefficient $D_{\text{eff}}$ depends on the specific model. 

We also find that the subleading contributions $A_0^{2k}(x,t)$ with $k>0$, to the large-time leading Gaussian behavior $A_0^0(x,t)$, generically satisfy an inhomogeneous diffusion equation of the form,
\bea
\left[\frac{\partial}{\partial t}-D_{\text{eff}}\frac{\partial^2 }{\partial x^2}\right]A_0^{2k}(x,t)=S_{2k}(x,t),
\label{strategy:inhomo_diff}
\eea
where the source term $S_{2k}(x,t)$ is determined by the lower order solutions $\{A_0^{2n}(x,t);\,n<k\}$. Therefore, starting from the Gaussian solution $A_0^0(x,t)$, the higher order contributions $A_0^{2k}(x,t)$ can be solved recursively for arbitrary $k$. 
Incidentally, \eref{g:diffeq} and \eref{g:diffinhomo} suggest a diffusive scaling ansatz, 
\bea
A_0^{2k}(x,t)=\frac{1}{t^k}\, q_{2k}\Big(\frac{x}{\sqrt{4D_\text{eff}t}}\Big)\,\frac 1{\sqrt{4\pi D_\text{eff}t}}\, \exp{\Big(-\frac{x^2}{4D_\text{eff}t}\Big)}.
\label{gen:a02k_ansatz}
\eea
Substituting the above ansatz in \eref{g:diffinhomo}, along with 
the scaling form 
\bea
S_{2k}(x,t)=\frac{1}{t^{k+1}}\, s_{2k}\Big(\frac{x}{\sqrt{4D_\text{eff}t}}\Big)\,\frac 1{\sqrt{4\pi D_\text{eff}t}}\, \exp{\Big(-\frac{x^2}{4D_\text{eff}t}\Big)},
\eea
yields an inhomogeneous Hermite differential equation for $q_{2k}(z)$ as,
\bea
q''_{2k}(z)-2z\,q'_{2k}(z)+4k\,q_{2k}(z)=s_{2k}(z).
\label{gen:hermite_inh}
\eea
The two solutions of the corresponding homogeneous Hermite differential equation \eref{gen:hermite_inh} are 
\begin{align}
U_{2k}(z)=H_{2k}(z)\qquad\text{and}\qquad V_{2k}(z)=z\, {}_1F_1\left(\frac 1 2-k,\frac 3 2,z^2\right).
\end{align}
where $H_{2k}(z)$ is the Hermite polynomial of order $2k$ and $_1F_1(a,b,z)$ is the confluent hypergeometric function. Note that, $U_{2k}(z)$ and $V_{2k}(z)$ are respectively even and odd functions of $z$. Therefore, remembering that $q_{2k}(z)$ must be an even function of $z$, the complete solution can be written as, 
\bea
\mspace{-15mu} q_{2k}(z)=C_{2k}\, U_{2k}(z) +\int_0^z dy\,\Big[ V_{2k}(z)\,U_{2k}(y)-U_{2k}(z)\,V_{2k}(y)\Big]\frac{s_{2k}(y)}{W_{2k}(y)},
\label{eq:qn_gen}
\eea 
where the Wronskian is given by (see~\ref{App:wronskian})
\bea
W_{2k}(y)=\begin{vmatrix}
U_{2k}(y)& V_{2k}(y)\\
U'_{2k}(y)& V'_{2k}(y)
\end{vmatrix}
=(-1)^k\frac{(2k)!}{k!}e^{y^2}.
\eea
The arbitray constant $C_{2k}$ in \eref{eq:qn_gen} is determined by neither the normalization of $\rho(x,t)$ nor any symmetries of $A_0^{2k}$. We take recourse to the moments of the distribution to determine $C_{2k}$. To this end, starting from the Fokker-Planck or master equation, we derive the moments $\la x^{2k}\ra/(4D_\rt t)^k$ and expand in powers of $\tau_0/t$ for $t\gg\tau_0$. 
 On the other hand, the same can be also obtained from the distribution \eref{gen:strategy} and \eref{gen:a02k_ansatz}. In particular, the coefficient of $(\tau_0/t)^k$ is given by $\int_{-\infty}^{\infty} dz~ e^{-z^2}  z^{2k}\, q_{2k}(z)/\sqrt{\pi}$ which involves $C_{2k}$. This can now be determined by comparing the coefficients of $(\tau_0/t)^k$ obtained by the two methods.

Note that, given the initial and boundary conditions, one, in principle, has all the information to find the complete solution of the Fokker-Planck equation.  Therefore, at a first glance, it might seem surprising that the moments are needed to determine the coefficients $C_{2k}$. However, only the boundary conditions are used to go from \eref{strategy:inhomo_diff} to \eref{eq:qn_gen}, and it is the moments, through which the initial condition is used, albeit in an unconventional way.

 In the following sections, we illustrate this framework with the examples of four well-known active particle models and calculate a few subleading contributions to the Gaussian distribution explicitly.

%%%%%%%%%%%%%%%%%%%%%%%%%%%%%%%

\section{Run and tumble particles in one dimension}\label{Sec:RTP}

Run-and-tumble dynamics, originally introduced as a model for bacterial motion, has emerged as one of the fundamental models for studying active motion.
An RTP runs with a constant velocity $v_0$ and changes its direction via intermittent tumbling events. The position $x(t)$ of a one-dimensional RTP evolves by the Langevin equation,
\bea
\dot{x}=v_0\,\sigma(t),
\label{1drtp:eq1}
\eea
where the dichotomous noise $\sigma(t)$ switches between $+1$ and $-1$ at a rate $\gamma$.
The corresponding Fokker-Planck equation for the probability density $P_{\sigma}(x,t)$ of finding the particle at position $x$ with a given state $\sigma$ at time $t$, is given by,
\bea
\frac{\partial P_\sigma}{\partial t}=-\sigma v_0 \frac{\partial P_\sigma}{\partial x}-\gamma P_\sigma+\gamma P_{-\sigma}\quad\textrm{ where }\sigma=\pm 1.
\label{rtp:fp1}
\eea
We are interested in the position distribution $\rho(x,t)= P_+(x,t)+P_-(x,t)$, which, from \eref{rtp:fp1}, follows 
\begin{subequations}
\bea
\frac{\partial \rho}{\partial t}&=&-v_0\frac{\partial Q}{\partial x},\label{1drtp:eqA}\\
\frac{\partial Q}{\partial t}&=&-v_0\frac{\partial \rho}{\partial x}-2\gamma Q,
\label{1drtp:eqB}
\eea
\end{subequations}
where $Q(x,t)= P_+(x,t)-P_-(x,t)$. Taking a derivative of \eref{1drtp:eqA} with respect to $t$ and using \eref{1drtp:eqB}, we have the telegrapher's equation,
\bea
\tau \frac{\partial ^2 \rho}{\partial t^2}
 + \frac{\partial \rho}{\partial t}=D_\rt\frac{\partial^2 \rho}{\partial x^2},
\label{eq:rtp_FPx}
\eea 
where $\tau=(2\gamma)^{-1}$ is the characteristic time of the RTP and $D_\rt=v_0^2/(2\gamma)$. Evidently, in the limit $\tau\to 0$ while keeping $D_\rt$ finite, the above equation reduces to the diffusion equation with the diffusion coefficient $D_\rt$. Equation \eref{eq:rtp_FPx} can be solved exactly for arbitrary $\tau$ to obtain the complete time-dependent solution $\rho(x,t)$, and hence, also the large-time Gaussian behavior~\cite{Malakar2017}. However, here our goal is to illustrate the perturbative procedure described in Sec.~\ref{Sec:genstrategy}, and we will use the exact solution to merely validate the same.

\subsection{Perturbative expansion of the Fokker-Planck equation}
We proceed to perturbatively solve \eref{eq:rtp_FPx}, treating $\tau$ as a small parameter and writing, 
\bea
\rho(x,t)=\sum_{k=0}^{\infty}\tau^k\,\rho_k(x,t),
\label{rtp:series}
\eea
as mentioned in \eref{gen:strategy}. Next we substitute the series \eref{rtp:series} in \eref{eq:rtp_FPx}, and obtain $\rho_k(x,t)$ order by order by equating coefficients of $\tau^k$. In particular,  we get the diffusion equation, 
 \bea
 \frac{\partial \rho_0}{\partial t}=D_\rt\frac{\partial^2 \rho_0}{\partial x^2},
 \eea
at the leading order, as announced in \eref{g:diffeq}, resulting in the Gaussian distribution,
\bea
\rho_0(x,t)=\frac{1}{\sqrt{4\pi D_\rt t}}\exp\left(-\frac{x^2}{4D_\rt t}\right),
\label{rtp:rho0}
\eea
where $D_\rt $ is identified as the diffusion coefficient.

In general, $\rho_k(x,t)$ for $k>0$ satisfies an inhomogeneous diffusion equation,
\bea
 \left[\frac{\partial }{\partial t}-D_\rt\frac{\partial^2 }{\partial x^2}\right] \rho_k= -\frac{\partial ^2 \rho_{k-1}}{\partial t^2},
 \label{1drtp:n}
 \eea
similar to the form announced in~\eref{strategy:inhomo_diff}, where the source term is obtained from the previous order solution $\rho_{k-1}(x,t)$. We substitute the ansatz [given in~\eref{gen:a02k_ansatz}],
\begin{align}
\rho_k(x,t)=\frac{1}{t^k}q_{2k}\left(\frac{x}{\sqrt{4D_\rt t}}\right)\frac{1}{\sqrt{4\pi  D_\rt t}}\exp\left(-\frac{x^2}{4 D_\rt t}\right),\label{rtp:ansatz}
\end{align}
 in \eref{1drtp:n} to find that $q_{2k}(z)$ satisfies an inhomogeneous Hermite differential equation,
\begin{align}
 q''_{2k}(z)-2z \,q'_{2k}(z)+4 k\, q_{2k}(z)&=s_{2k}(z),
\label{rtp:inhomo}
\end{align}
as announced in \eref{gen:hermite_inh}. Here the source term is given by,
\begin{equation}
s_{2k}(z)=
z^2q''_{2k-2}(z)+z \left(1+4k- 4z^2\right) q'_{2k-2}(z)
-\left[1+4z^2-4(z^2-k)^2\right] q_{2k-2}(z). \label{eq:fn}
\end{equation} 
Equation \eqref{rtp:inhomo} is augmented by the boundary condition $q_{2k}(z)\exp(-z^2)\to 0$ as $z\to\pm\infty$ and $q_0(z)=1$. Moreover, for the initial condition $\sigma(0)=\pm 1$ with equal probability $1/2$, we must have $q_{2k}(-z)=q_{2k}(z)$ for all $k$. The general solution, respecting this symmetry,  is given by \eref{eq:qn_gen}  for $k\geq 1$, with the arbitrary constant $C_{2k}$ yet to be determined. 

By multiplying \eref{rtp:inhomo} with $e^{-z^2}$ and integrating over $z$, we find that 
\begin{align}
\int_{-\infty}^{\infty}dz \,e^{-z^2}q_{2k}(z)=(k-1)\int_{-\infty}^{\infty}dz \,e^{-z^2}q_{2k-2}(z).
\end{align}
Therefore, $\int_{-\infty}^{\infty}dz \,e^{-z^2}q_{2k}(z)=0$ for $k\geq 1$. However, this condition cannot determine $C_{2k}$ in \eref{eq:qn_gen}, as $\int_{-\infty}^{\infty}dz \,e^{-z^2}\,U_{2k}(z)=0$. Therefore, we use a different procedure involving the moments, as described below.

\subsection{Moments }
From the Fokker-Planck equation  \eqref{eq:rtp_FPx}, the $2k$-th  moment $M_{2k}(t)= \la x^{2k}(t)\ra$ follows the differential equation,
\begin{equation}
\tau \frac{d^2 M_{2k}}{d t^2}+ \frac{d M_{2k}}{d t}=  2k(2k-1) D_\rt M_{2k-2}, \quad \text{with} \quad  M_{0}(t)=1. \label{eq:M_recur}
\end{equation}
Since $\sigma(t)$ in \eref{1drtp:eq1} remains unchanged for $t\ll\tau$, we have $M_{2k}(t) = (v_0t)^{2k}+o(t^{2k})$ as $t \to 0$. Taking a Laplace transform $\widetilde{M}_{2k}(s)=\int_{0}^{\infty}e^{-st}M_{2k}(t) dt$ in \eref{eq:M_recur} and using the initial conditions $M_{2k}(0)=0=M'_{2k}(0)$ for $k\geq 1$, we get,
\bea
\widetilde{M}_{2k}(s)=\frac{2k(2k-1)D_\rt}{s(1+\tau s)}\,\widetilde{M}_{2k-2}(s)=\frac{(2k)!\,D_\rt^k}{s^{k+1}(1+\tau s)^k},
\eea
where we have used $\widetilde{M}_{2k}(0)=1/s$. This can be inverted to get the moments exactly,
\bea
\fl  M_{2k}(t)=(4 D_\rt t)^k \,\frac{1}{2}\,\Gamma \left(k+\frac{1}{2}\right)  \,\sqrt{\frac{t}{\tau}}\,\exp\left(-\frac{t}{2\tau}\right)\left[I_{k+\frac{1}{2}}\left(\frac{t}{2 \tau }\right)+I_{k-\frac{1}{2}}\left(\frac{t}{2 \tau }\right)\right],
\label{rtp:exactmoment}
\eea
where $I_{\nu}(z)$ is a modified Bessel function of first kind.

\subsection{Position distribution}
We are now in a position to obtain $\rho_{2k}(x,t)$ explicitly by determining the constants $C_{2k}$. The $2k$th moment obtained from the distribution \eref{rtp:series} and \eref{rtp:ansatz} is given by,
\bea
\frac{M_{2k}(t)}{(4D_\rt t)^k}=\sum_{n=0}^k \left(\frac{\tau}t \right)^n \int_{-\infty}^{\infty} \frac{dz}{\sqrt{\pi}}~ e^{-z^2}  z^{2k}\, q_{2n}(z),
\label{rtpmomentcompare}
\eea
where we have used relation \eref{app_b:rtp}. The integrals in the above equation can be evaluated using the general solution of $q_{2n}(z)$ given by \eref{eq:qn_gen} in terms of the unknown constants $C_{2n}$. On the other hand, we can expand $M_{2k}(t)/(4D_\rt t)^k$ from \eref{rtp:exactmoment}, as a series in powers of $\tau/t$ for $t\gg\tau$. We determine the constant $C_{2k}$ by comparing the coefficients of $(\tau/t)^k$ in the expansion of the moments obtained by the two methods.

For example, setting $k=1$ in \eref{eq:fn} we have, $s_2(z) = 3-12z^2 + 4z^4,$ using which in \eref{eq:qn_gen} leads to,
\bea
q_2(z) =  C_2 H_2(z) + \frac 32z^2  - z^4 .\label{rtp:q2_sol}
\eea
Therefore, the coefficients of $\tau/t$ in \eref{rtpmomentcompare} is given by,
\bea 
\int_{-\infty}^{\infty} \frac{dz}{\sqrt{\pi}}~ e^{-z^2}  z^2\, q_2(z) =2C_2- \frac 34.
\label{rtp_moments_tau/t}
\eea
On the other hand from \eref{rtp:exactmoment}, we have
\bea
\frac{M_2(t)}{4 D_\rt t} = \frac 12 - \frac {\tau}{2t}  + O(e^{-t/\tau}). \label{eq:m2}
\eea
where the coefficient of $\tau/t$ is $-1/2$. Comparing it with \eref{rtp_moments_tau/t}, we get $C_2=1/8$. This gives,
\bea
q_2(z)=- \frac 14 (1-8z^2 + 4 z^4).\label{rtp:q2}
\eea

For the next order contribution, we set $k=2$ in \eref{eq:fn} to get $s_4(z) = -\frac{15}{4} + 75 z^2 - 120 z^4 + 44 z^6 - 4 z^8$. Using this in \eref{eq:qn_gen} leads to,
\bea
q_4(z) = C_4 H_4(z)-\frac{1}{8} z^2 \left(15-55z^2+32z^4-4 z^6\right) .\label{rtp:q4_sol}
\eea
On the other hand from \eref{rtp:exactmoment}, we have
\bea
\frac{M_4(t)}{(4 D_\rt t)^2} = \frac{3}{4}-\frac{3}{2}\left(2+ e^{-\frac{t}{\tau }}\right)\frac{\tau}{t}+\frac{9}{2} \left(1- e^{-\frac{t}{\tau }}\right)\left(\frac{\tau}{t}\right)^2. \label{eq:m4}
\eea
Comparing the coefficients of $(\tau/t)^2$ in the above equation  and that obtained from \eref{rtpmomentcompare}, we have $C_4=-1/128$, which leads to,
\bea
q_4(z)=-\frac{1}{32}\left(3 + 48 z^2 - 216 z^4 + 128 z^6 - 16 z^8\right).\label{rtp:q4}
\eea
Proceeding similarly by setting $k=3$ in \eref{eq:fn} we get $s_6(z) = -\frac{105}{32}-\frac{735}{8}z^2+\frac{5775}{8}z^4-875z^6+\frac{665}{2}z^2-46 z^{10}+2z^{12}$. Using this in \eref{eq:qn_gen} leads to the general solution for $q_6(z)$,
\bea
\fl \quad q_6(z) = C_6 H_6(z)-\frac{z^2}{192} (315 + 1260 z^2 - 4788 z^4 + 3000 z^6 - 576 z^8 + 32 z^{10}).\label{rtp:q6_sol}
\eea
Comparing the coefficient of $(\tau/t)^3$ in $M_6(t)/(4D_\rt t)^3$ from \eref{rtp:exactmoment} and \eref{rtpmomentcompare}, we get $C_6=1/1024$. Using this in \eref{rtp:q6_sol}, we get,
\bea
\fl q_6(z)=-\frac{1}{384}\left(45 + 360 z^2 + 2700 z^4 - 9600 z^6 + 6000 z^8 - 1152 z^{10} + 64 z^{12}\right).\label{rtp:q6}
\eea

We can go on to find the higher  order contributions to the position distribution following the same procedure. The sub-leading contributions $q_2(z),$ $q_4(z),$ and $q_6(z)$  can also be extracted from the exact solution and match exactly with those obtained here (see \ref{app:rtp_exact}).

%%%%%%%%%%%%%%%%%%
\section{Active Ornstein-Uhlenbeck particles}\label{Sec:AOUP}
Active Ornstein-Uhlenbeck particle (AOUP) is perhaps the simplest mathematical model for active motion. The position of an AOUP, in the absence of any potential, undergoes an  overdamped motion,
\bea
\dot{x}(t)=v(t),
\label{under:langevin1}
\eea
where the self-propulsion velocity $v(t)$ follows an Ornstein-Uhlenbeck process,
\bea
\tau\dot{v}(t)=- v(t)+\sqrt{2D}\,\eta (t).
\label{under:langevin2}
\eea
Here $\eta(t)$ is a white noise with $\la\eta(t)\eta(t')\ra=\delta(t-t')$. The persistence time $\tau$ and the noise strength $D$ determine the strength of the propulsion velocity as, $\la v(t)v(t')\ra=(D/\tau)\, e^{-|t-t'|/\tau}$ for $t,\,t'\gg\tau$.

The Fokker-Planck equation governing the joint probability distribution $P(x,v,t)$ is given by,
\bea
\frac{\partial P}{\partial t}=-v\frac{\partial P}{\partial x}+\frac 1{\tau} \frac{\partial }{\partial v}(vP)+\frac D{\tau^2}\,\frac{\partial ^2 P}{\partial v^2}.
\label{aoup_FP1}
\eea
We consider the initial condition $x=0$, $v=0$ at $t=0$. At time-scales much larger than the persistence time $\tau$, we expect the AOUP to display a diffusive behavior with an effective diffusion constant $D$. Therefore, to study the approach to the diffusive regime, it is natural use the scaled variables $u=v\sqrt{\tau/D}$ and $y= x/\sqrt{D}$. Consequently,  \eref{aoup_FP1} becomes ,
\bea
\varepsilon^2 \frac{\partial P}{\partial t}=-\varepsilon u \frac{\partial P}{\partial y}+\mathcal{L}_u P\quad \text{with}~~ \mathcal{L}_u=\frac{\partial}{\partial u}u+\frac{\partial ^2}{\partial u^2}\quad \text{and  }\varepsilon=\sqrt \tau.
\label{e:underdampedfp}
\eea

The isotropic initial condition ensures that the marginal position distribution $\rho(y,t)=\int_{-\infty}^{\infty}du P(y,u,t)$ is symmetric at all times. This implies that all the odd moments of the position vanish. Moreover, since \eref{e:underdampedfp} remains invariant under the transformation $(y,\varepsilon)\to(-y,-\varepsilon)$, $\rho(y,t)$ contains only even powers of $\varepsilon$. As mentioned in the general perturbative strategy, we need the $2n$-th moment $\la y^{2n}(t) \ra$ to determine the distribution uniquely at order $(\varepsilon^2/t)^n.$ In the following subsection we describe the technique of computing the moments recursively.

\subsection{Moments} \label{aoup:moments}
Since the position distribution of an AOUP is Gaussian, all the moments can be found easily (see \ref{App:aoup}). However, here our aim is to demonstrate the recursive procedure for obtaining the moments without making use of the exact known distribution.
 To this end, let us first define the correlation function,
\begin{align}
M(k,n,t)=\la y^k u^n \ra=\int_{-\infty}^{\infty}dy\int_{-\infty}^{\infty}du~y^k u^n P(y,u,t).
\end{align} 
 Note that, $M(k,0,t)=\la y^{k}(t) \ra$ denote the position moments. Multiplying both sides of \eref{e:underdampedfp} and then integrating over $y$ and $u$, we obtain, for $n,k\geq 1$,
 \begin{align}
\left[ \varepsilon^2\frac{d}{dt}+n\right]M(k,n,t)=\varepsilon k M(k-1,n+1,t)+n(n-1)M(k,n-2,t).
 \end{align}
This is a first order ODE in time with the general solution,
\begin{align}
M(k,n,t)=\frac{1}{\varepsilon^2}\int_{0}^{t}dt' e^{-\frac{(t-t')n}{\varepsilon^2}}\left[\varepsilon k M(k-1,n+1,t')+n(n-1)M(k,n-2,t')\right],
\label{under:moments}
\end{align}
where we have used the initial condition $M(k,n,0)=0$. Moreover, the normalization condition and the vanishing mean of the Gaussian white noise $\la\eta(t)\ra=0$  lead to 
\begin{align}
M(0,0,t)=1 \quad\text{ and }\quad M(2k+1,0,t)=M(0,2k+1,t)=0,
\label{eq:aoup_moments}
\end{align}
respectively. The position moments, i.e., $n=0$ and $k\geq 1$ are thus given by,
\begin{align}
M(k,0,t)=\frac{k}{\varepsilon}\int_{0}^{t}dt' M(k-1,1,t'). 
\label{aoup_posmom}
\end{align}
 
%%%%%%%%%%%%%%%%%%%%%%%%%%% 
\begin{figure}
\centering\includegraphics[width=0.7 \hsize]{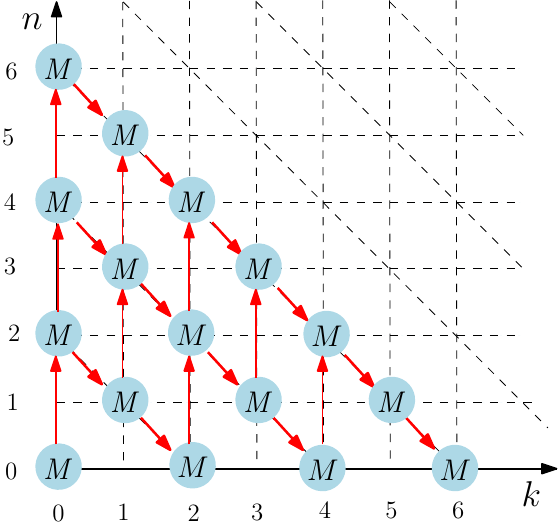}
\caption{The illustration of the recursive connections between the correlation function $M(k,n,t)$ for different $k,n$ with even $k+n$; see equation \eref{under:moments}.}
\label{f:aoupmoments}
\end{figure}
%%%%%%%%%%%%%%%%%%%%%%%%%%%

It follows from the structure of \eref{under:moments} and \eref{eq:aoup_moments} that the correlation functions $\{M(k,n,t)\}$ for different values of $(k,n)$ are non-zero only on the even $k+n$ sub-lattice. The network formed by non-zero $M(k,n,t)$ for different $k,n$ with even $(k+n)$ is illustrated in Fig.~\ref{f:aoupmoments}.

We now illustrate the recursive procedure by computing the variance  explicitly, for which we need $M(1,1,t)$ [putting $k=2$ in \eref{aoup_posmom}], which in turn, depends on $M(0,2,t)$, which further depends on $M(0,0,t)=1$ [using \eref{eq:aoup_moments}]. Following this route, we first obtain $M(0,2,t)$,
\begin{align}
M(0,2,t)=1-e^{-\frac{2 t}{\varepsilon ^2}},
\end{align}
which can be used in \eref{eq:aoup_moments} for $k=n=1$ to get,
\begin{align}
M(1,1,t)=\varepsilon\left(1+e^{-\frac{2t}{\varepsilon^2}}-2e^{-\frac{t}{\varepsilon^2}}\right).
\end{align}
Finally, the variance comes out to be,
\begin{align}
M(2,0,t)=2t-\varepsilon^2 \left(3-4e^{-\frac{t}{\varepsilon^2}}+e^{-\frac{2t}{\varepsilon^2}}\right).
\label{under:m2}
\end{align}
For the fourth order position moment $M(4,0,t)$, we need to compute the correlation functions $M(0,2,t),~M(2,2,t)$, and $M(3,1,t)$, in addition to what we already had during the evaluation of $M(2,0,t).$ In general, to compute $M(2j-2,0,t)$ requires all the non-zero correlation functions in the region bounded by $k=0$, $n=0$ and $k+n=2j-2$. Therefore, the subsequent computation of $M(2j,0,t)$ requires the computation of the correlation functions $M(k,n,t)$ along the line $k+n=2j$ starting with $M(0,2j,t)$. This network with the recursive connections is illustrated in \fref{f:aoupmoments} by red arrows. Following this procedure, we evaluate the fourth and sixth position moments as,
\begin{align}
M(4,0,t)=12t^2-12\varepsilon^2t\left(3-4e^{-\frac{t}{\varepsilon^2}}+e^{-\frac{2t}{\varepsilon^2}}\right)+3\epsilon^4\left(3-4e^{-\frac{t}{\varepsilon^2}}+e^{-\frac{2t}{\varepsilon^2}}\right)^2,
\label{under:m40}
\end{align}
and
\begin{align}
M(6,0,t)=120t^3-180&\varepsilon^2 t^2\left(3-4e^{-\frac{t}{\varepsilon^2}}+e^{-\frac{2t}{\varepsilon^2}} \right) +90 \varepsilon^4 t\left(3-4e^{-\frac{t}{\varepsilon^2}}+e^{-\frac{2t}{\varepsilon^2}}\right)^2\cr
&-15\varepsilon^6\left(3-4e^{-\frac{t}{\varepsilon^2}}+e^{-\frac{2t}{\varepsilon^2}}\right)^3.
\end{align}
These expressions for the moments, of course, match with the exact moments of the Gaussian marginal distribution given in \ref{App:aoup}. We will use these position moments to determine the position distribution perturbatively in $\varepsilon$.

\subsection{Position Distribution}
To obtain the position distribution perturbatively, it is important to first note that the distribution of $u$, evolving by the Fokker-Planck operator $\mathcal{L}_u$, corresponding to the Ornstein-Uhlenbeck process, reaches a steady state $\psi_0(u)$. In fact, 
the eigenvalues of $\mathcal{L}_u$ are given by $-n$ with $n=0,1,2,\dotsc,\infty$. The corresponding eigenfunctions satisfying $\mathcal{L}_u \psi_n(u)=-n\psi_n(u)$, are given by,
  \bea
 \psi_n(u)=\frac{e^{-u^2/2}H_n(u/\sqrt{2})}{\sqrt{2\pi}2^n n!},
 \label{hermiteeigenvalue}
 \eea
where $H_n(z)$ is the Hermite polynomial of order $n$. From orthonormality relations of Hermite polynomials, it follows that,
\bea
\int_{-\infty}^{\infty}du\, \psi_n(u)\,H_m\left(\frac u{\sqrt{2}}\right)=\delta_{m,n}.
\label{hermite_orthonormal}
\eea
Thus, we can always express the solution of \eref{e:underdampedfp} in the eigenbasis of $\mathcal{L}_u$ as,
 \bea
 P(y,u,t)=\sum_{n=0}^{\infty} \psi_n(u)F_n(y,t),
% \sum_{n=0}^{\infty} \frac{e^{-u^2/2}H_n(u/\sqrt{2})A_n(y,t)}{\sqrt{2\pi}2^n n!}=
 \label{hnbasis}
 \eea 
 Integrating \eref{hnbasis} with respect to $y$ gives the marginal distribution of $u$,
\bea
\int_{-\infty}^\infty P(y,u,t)\,dy= \sum_{n=0}^{\infty} b_n\,\psi_n(u) \,e^{-nt/\varepsilon^2},
\label{under:p_ut}
\eea
where $b_0=1$ and $b_n$, for $n>0$ are determined by the initial condition. When the initial value of $u$ is drawn from the stationary distribution $\psi_0(u)$, the marginal distribution of $u$ does not evolve, i.e., $\int_{-\infty}^\infty P(y,u,t)\,dy= \psi_0(u)$ at all times, i.e., $b_n=0$ for $n>0$.

Our goal is to obtain the marginal position  distribution,
\bea
\rho(y,t)\equiv\int_{-\infty}^{\infty} P(y,u,t)\,du=F_0(y,t),
\eea
 where the second equality follows from \eref{hnbasis} and \eref{hermite_orthonormal}. In general, the function $F_n(y,t)$ is formally given by,
\bea
F_n(y,t)=\int_{-\infty}^{\infty} P(y,u,t)\,H_n\left(\frac{u}{\sqrt{2}}\right)\,du.
\label{under:F_n}
\eea
However, this equation cannot be used to determine $F_n(y,t)$ as $P(y,u,t)$ is unknown.

We proceed by deriving a differential equation for the time evolution of $F_n(y,t)$ from \eref{e:underdampedfp}. Substituting  \eref{hnbasis} in \eref{e:underdampedfp} and using \eref{hermiteeigenvalue}, we get,
 \begin{align}
\varepsilon^2 \sum_{n=0}^{\infty} \psi_n(u)\frac{\partial F_n(y,t) }{\partial t}=-\varepsilon u\sum_{n=0}^{\infty} \psi_n(u)\frac{\partial  F_n(y,t)}{\partial y}-n\sum_{n=0}^{\infty} \psi_n(u)F_n(y,t).
 \end{align}
  Multiplying both sides by $H_m(u/\sqrt{2})$ and integrating over $u$, we get, 
 \bea
 \left[\varepsilon^2\frac{\partial }{\partial t}+m\right]F_m=-\frac{\varepsilon}{\sqrt{2}}\left(2m\frac{\partial F_{m-1}}{\partial y}+\frac{\partial F_{m+1}}{\partial y}\right),
\label{under:am}
 \eea
where we have used the orthonormality condtion \eref{hermite_orthonormal} and the identity,
$u\psi_n(u)=\bigl[\psi_{n+1}(u)+2n\psi_{n-1}(u)\bigr]/\sqrt{2}$.

From \eref{hnbasis} and \eref{under:p_ut}, we must have 
\bea
\int_{-\infty}^{\infty}F_m(y,t)\,dy=b_m \,\nu^m \quad\text{with }\nu=e^{-t/\varepsilon^2}.
\label{am(yt)}
\eea
Since $\nu$ cannot have a Taylor series expansion around $\varepsilon=0$ because of the essential singularity, $F_m(y,t)$ can have a general power series expansion 
\bea
F_m(y,t)=\sum_{l=0}^{\infty}\nu^l F_{m,l}(y,t).
\label{u:aml}
\eea
where $F_{m,l}(y,t)$ is analytic and can be again expanded in Taylor series around $\varepsilon=0$. It follows from \eref{am(yt)} and \eref{u:aml} that, $\int_{-\infty}^{\infty} F_{m,l}(y,t)\,dy=b_m\delta_{m,l}$. By substituting \eref{u:aml} in \eref{under:am}, it easily seen that, $F_{m,l}(y,t)$ satisfies the differential equation,
\bea
 \left[\varepsilon^2\frac{\partial}{\partial t}+(m-l)\right]F_{m,l}=-\frac{\varepsilon}{\sqrt{2}}\left(2m\frac{\partial F_{m-1,l}}{\partial y}+\frac{\partial F_{m+1,l}}{\partial y}\right).
\label{under:aml}
 \eea
For $t\gg\varepsilon^2$, it suffices to consider only the first term of the series in \eref{u:aml}, as the $l\geq 1$ terms decay exponentially, i.e., $F_{m}(y,t)= F_{m,0}(y,t)+O(e^{-t/\varepsilon^2})$. Note that, $F_{m}(y,t)= F_{m,0}(y,t)$ for all $t$, when the initial values of $u$ are drawn from $\psi_0(u)$.

Now, we proceed to determine $F_{m,0}(y,t)$ by solving \eref{under:aml} with $l=0$ perturbatively. To this end, we write,
\bea
F_{m,0}(y,t)=\sum_{k=0}^{\infty} \varepsilon^k A_m^k(y,t).
\label{under:akm}
\eea
Putting \eref{under:akm} in \eref{under:aml} (with $l=0$) and collecting terms of the order $\varepsilon^k$, we get,
 \bea
 \frac{\partial A_m^{k-2}}{\partial t}=-\frac{1}{\sqrt{2}}\frac{\partial }{\partial y}\left[2m A_{m-1}^{k-1}+ A_{m+1}^{k-1}\right]-m A_m^k,
 \label{under:recur}
 \eea
 with $A_{m}^k=0$ for $k<0$. In the following, we show that $A_m^k$ in \eref{under:akm} is non-zero only when $(m+k)$ is even. We start by putting $k=0$ in \eref{under:recur}, which gives $A_{m}^0=\delta_{m,0}A_0^0$. Next putting $k=1$, we get
\begin{equation}
A_m^1=-\delta_{m,1}\sqrt{2}\frac{\partial A_0^0}{\partial y}+\delta_{m,0}\,A_0^1.
\end{equation}
Note that, $A_0^1=0$; in fact, $A_0^{2k+1}=0$ for all $k$. This is because the marginal distribution $A_0(y,t)$ is symmetric in $y$ and \eref{under:am} has the symmetry $(y,\varepsilon)\to(-y,-\varepsilon)$.

%%%%%%%%%%%%%%%%%
\begin{figure}
\centering\includegraphics[width=0.7\hsize]{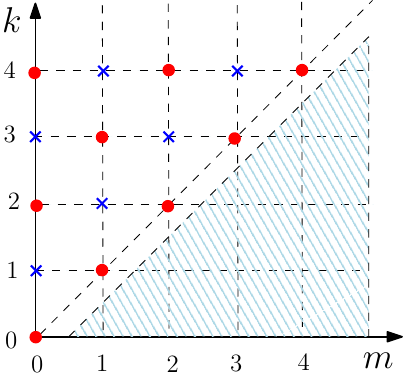}
\caption{Graphical representation of recursive determination of  $A_m^k$, given by \eref{under:recur}. The red dots represent the non-zero $A_m^k$, whereas the blue cross represents $A_m^k=0$.}
\label{underdamped_am}
\end{figure}
%%%%%%%%%%%%%%

Now, we can systematically proceed by putting $k=2,3, 4,\dotsc$. This process is best illustrated graphically  on the $(m,k)$ plane (see Fig.~\ref{underdamped_am}). The fact that $A_0^0\neq 0$ (yet to be determined) and  $A_0^1=0$, leads to $A_m^k=0$ for odd $m+k$ and $m >k$. Therefore, \eref{under:akm} can be refined to,
 \begin{align}
F_{m,0}(y,t)&=\sum_{k=m}^{\infty}\varepsilon^{k}A^{k}_m(y,t)\quad\text{with  } (m+k)\text{ even},
\label{under:amk1}
\end{align}
In particular, the  marginal distribution of $y$ is given by,
\bea
\rho(y,t)=F_{0,0}(y,t)+O(e^{-t/\varepsilon^2})\quad \text{where }F_{0,0}(y,t)=\sum_{k=0}^{\infty}\varepsilon^{2k}A^{2k}_0(y,t).
\label{marginal:correction}
\eea 
as stated in \eref{gen:strategy}.
We proceed to compute $A_0^{2k}(y,t)$ recursively, which from 
\eref{under:recur} satisfies the differential equation,
\bea
\frac{\partial A_0^{2k}}{\partial t}=-\frac{1}{\sqrt{2}}\frac{\partial A_{1}^{2k+1}}{\partial y}.
\label{under:a02k}
\eea
For $k=0$, we have,
%Therefore, the equation for $A_0^0(y,t)$ involves $A_1^1(y,t)$,
\bea
 \frac{\partial A_{0}^{0}}{\partial t}=-\frac{1}{\sqrt{2}}\frac{\partial A_{1}^{1}}{\partial y}.
 \label{a00_1}
 \eea
To obtain a closed form equation for $A_0^0(y,t)$ we need $A_1^1(y,t)$ in terms of  $A_0^0(y,t)$. To this end, we note that, in general, $A_m^m(y,t)$ is related to $A_0^0(y,t)$ via a simple relation [from \eref{under:recur}], 
 \bea
 A_m^m=-\sqrt{2}\frac{\partial A_{m-1}^{m-1}}{\partial y}=(-\sqrt{2})^m\frac{\partial^m A_{0}^{0}}{\partial y^m}.
 \label{under:am_m}
 \eea
Substituting $A_1^1(y,t)$ from \eref{under:am_m} in \eref{a00_1}, yields the diffusion equation,
 \bea
 \frac{\partial A_{0}^{0}}{\partial t}=\frac{\partial ^2 A_{0}^{0}}{\partial y^2},
 \label{diffusiona00}
 \eea 
 as stated in \eref{g:diffeq}.
  Thus the normalized marginal distribution to order $\varepsilon^0$ is given by,
\bea
A_0^0(y,t)=\frac{1}{\sqrt{4\pi t}}\exp\left(-\frac{y^2}{4t}\right).
 \eea 
 Applying $(-\sqrt{2})^m(\partial ^m/\partial y^m)$ on both sides of \eref{diffusiona00} and using \eref{under:am_m}, we see that $A_m^m(y,t)$ also satisfies the diffusion equation,
\bea
 \frac{\partial A_{m}^{m}}{\partial t}=\frac{\partial ^2 A_{m}^{m}}{\partial y^2}.
 \label{diffusionamm}
 \eea 
 For the next order correction $A_0^2(y,t)$ to the marginal distribution $\rho(y,t)$, from \eref{under:a02k} we get,
 \bea
 \frac{\partial A_{0}^{2}}{\partial t}=-\frac{1}{\sqrt{2}}\frac{\partial A_{1}^{3}}{\partial y}.
 \label{u:a02}
 \eea
In fact, we find from \eref{under:recur} that in general $A_m^{m+2}(y,t)$ is given by,
\bea
 A_m^{m+2}&=&-\sqrt{2}\frac{\partial A_{m-1}^{m+1}}{\partial y}-\frac{1}{m}\left(\frac{\partial A_m^m}{\partial t}+\frac{1}{\sqrt{2}}\frac{\partial A_{m+1}^{m+1}}{\partial y}\right).
\eea
Following \eref{under:am_m} and \eref{diffusionamm}, the terms in the parenthesis cancel each other resulting in,
\bea
A_m^{m+2}=-\sqrt{2}\frac{\partial A_{m-1}^{m+1}}{\partial y}=(-\sqrt{2})^m\frac{\partial^m A_{0}^{2}}{\partial y^m}.
\label{u:am-m+2:1}
\eea
 In particular, using  $A_1^3 = -\sqrt{2}(\partial A_{0}^{2}/\partial y)$ in \eref{u:a02}, we find that $A_0^2(y,t)$ also satisfies a diffusion equation,
\bea
 \frac{\partial A_{0}^{2}}{\partial t}=\frac{\partial ^2 A_{0}^{2}}{\partial y^2}.
 \eea
As a consequence, similar to \eref{diffusionamm}, $A_m^{m+2}(y,t)$ also satisfies a diffusion equation,
\bea 
\frac{\partial A_{m}^{m+2}}{\partial t}=\frac{\partial ^2 A_{m}^{m+2}}{\partial y^2}.
\label{u:am-m+2:2}
\eea
 
 To see a general pattern emerging, let us evaluate the $O(\varepsilon^4)$ correction $A_0^4(y,t)$. From \eref{under:a02k}, it satisfies,
 \bea
 \frac{\partial A_0^{4}}{\partial t}=-\frac{1}{\sqrt{2}}\frac{\partial A_{1}^{5}}{\partial y}.
 \eea 
On the other hand, from \eqref{under:recur}, 
\bea
A_m^{m+4}=-\sqrt{2}\frac{\partial A_{m-1}^{m+3}}{\partial y}-\frac{1}{m}\left(\frac{\partial A_m^{m+2}}{\partial t}+\frac{1}{\sqrt{2}}\frac{\partial A_{m+1}^{m+3}}{\partial y}\right).
\eea
Following \eref{u:am-m+2:1} and \eref{u:am-m+2:2}, again the expression in the parenthesis vanishes resulting in,
\bea
A_m^{m+4}=-\sqrt{2}\frac{\partial A_{m-1}^{m+3}}{\partial y}=(-\sqrt{2})^m\frac{\partial^m A_{0}^{4}}{\partial y^m}.
\eea

In general, for a given set $(m,\,k)$, if
\bea
\fl\qquad A_m^{m+2(k-1)}=-\sqrt{2}\frac{\partial A_{m-1}^{m-1+2(k-1)}}{\partial y}\quad\text{and}\quad
\frac{\partial A_{m}^{m+2(k-1)}}{\partial t}=\frac{\partial ^2 A_{m}^{m+2(k-1)}}{\partial y^2}
\eea
hold, then it can be shown that, in the next order,
\bea
\fl \qquad A_m^{m+2k}=(-\sqrt{2})\frac{\partial A_{m-1}^{m-1+2k}}{\partial y}\quad\text{and}\quad
\frac{\partial A_{m}^{m+2k}}{\partial t}=\frac{\partial ^2 A_{m}^{m+2k}}{\partial y^2}
\label{diff:amk}
\eea
are satisfied. The validity of this recursive procedure has already been illustrated explicitly for arbitrary $m$ and $k=0,1,2$. Therefore, by induction, the relations given in \eref{diff:amk} are valid for any $(m,k)$. In particular, for $m=0$, we have,
\bea
\frac{\partial A_{0}^{2k}}{\partial t}=\frac{\partial ^2 A_{0}^{2k}}{\partial y^2},
\label{u:a02kgeneral}
\eea
as stated in \eref{strategy:inhomo_diff} with the inhomogeneous part $S_{2k}(y,t)=0$.
Even though $A_0^{2k}(y,t)$ for $k\geq 1$ satisfies the same diffusion equation as $A_0^{0}(y,t)$, the solution of \eref{u:a02kgeneral} for $k\geq 1$ cannot be a simple Gaussian, 
since the normalization of $\rho(y,t)$ demands that 
\bea
\int_{-\infty}^{\infty}A_0^{2k}(y,t)\,dy=\delta_{k,0}.
\label{normalizationa02k}
\eea
 Nevertheless, because of the diffusive scaling, we anticipate a solution of \eref{u:a02kgeneral} of the form (that will be shown to hold {\it a posteriori}),
\bea
A_0^{2k}(y,t)=\frac{1}{t^{k}}\,q_{2k}\left(\frac{y}{\sqrt{4t}}\right)\frac{e^{-y^2/(4t)}}{\sqrt{4\pi t}}.
\label{under:ansatz}
\eea 
 Substituting \eref{under:ansatz} in \eref{u:a02kgeneral}, we see $q_{2k}(z)$ satisfies the Hermite differential equation,
 \bea
 q_{2k}''(z)-2z\,q'_{2k}(z)+4k\,q_{2k}(z)=0,
 \eea 
as mentioned in \eqref{gen:hermite_inh} with the inhomogeneous part $s_{2k}(z)=0$. Therefore, using the fact that $q_{2k}(z)$ is an even function of $z$, we get
 \bea
 q_{2k}(z)=C_{2k} H_{2k}(z),
 \label{under:q2m}
 \eea
 where $H_{2k}(z)$ is the $2k$-th degree Hermite polynomial and the constant $C_{2k}$ is yet to be determined for each $k\geq 1$. Note that, $H_0(z)=1$ and $C_0=1$.
 
  The condition \eref{normalizationa02k} is trivially satisfied for any arbitrary sets of $\{C_{2k}\}$s as $\int_{-\infty}^{\infty}dz\,e^{-z^2}H_{2k}(z)=\sqrt{\pi}\,\delta_{k,0}$. Therefore, to determine $C_{2k}$ we take recourse to position moments $M(2k,0,t)$ defined in in Sec.~\ref{aoup:moments}.
  
 From \eref{marginal:correction}, \eref{under:ansatz} and \eref{under:q2m}, the position moment $M(2k,0,t)$ can be expressed as,
 \bea
 \fl\qquad
\frac{M(2k,0,t)}{(4 t)^{k}}&=& \sum_{m=0}^{k}\left(\frac{\varepsilon^2}{t}\right)^{m}C_{2m}\int_{-\infty}^{\infty}\frac{dz}{\sqrt{\pi}}\,z^{2k}\,H_{2m}(z)\,e^{-z^2}+O(e^{-t/\varepsilon^2}),
\label{aoup:momentcomparison}
\eea
where we have used the fact that $\int_{-\infty}^{\infty}dz\,z^{2k}\,H_{2m}(z)\,e^{-z^2}=0$ for $m>k$ because of the orthogonality of Hermite polynomials as $z^{2k}$ can be only expressed as a linear combination of Hermite polynomials of degree $2k$ and lower. In fact, for $m\leq k$, the integral in \eref{aoup:momentcomparison} can be explicitly evaluated as,
\begin{align}
\int_{-\infty}^{\infty}\frac{dz}{\sqrt{\pi}}\,z^{2k}\,H_{2m}(z)\,e^{-z^2}=\frac{(2m)!\,(2k)!}{2^{2k-\frac12}\, m!\,k!}\,(-1)^m \,{}_2 F_1\left(-m,k+\frac 12,\frac 12,1\right).
\end{align}
In Sec.~\ref{aoup:moments}, $M(2k,0,t)$ has been calcualted exactly, from which the left hand side of \eref{aoup:momentcomparison} becomes a polynomial of degree $k$ in $(\varepsilon^2/t)$  for large $t$. Therefore, comparing the coefficients of $(\varepsilon^2/t)^m$ on both sides yields $C_{2m}$ for $m\leq k$. For example, the first three non-zero constants are given by $C_2=-3/8$, $C_4=9/128$ and $C_6=-9/1024$.

The validity of this perturbative procedure discussed above can be explicitly checked for the AOUP, since it is a Gaussian process. The joint distribution $P(x,v,t)$, which is a bivariate Gaussian, is completely determined by the covariance matrix. In ~\ref{App:aoup}, we explicitly show that the results obtained from the perturbative analysis are identical to those obtained from the exact joint distribution.

\section{Active Brownian particles}\label{Sec:ABP}
Active Brownian particle (ABP) is often used to model the trajectories of a large class of Janus particles and various other micro-swimmers. The self-propulsion direction of an ABP undergoes a rotational diffusion while keeping the speed constant.
The Langevin equation for the position $(x,y)$ of an ABP in two dimensions is given by,
\begin{align}
\dot{x}&=v_0 \cos\theta(t),\quad \dot{y}=v_0 \sin\theta(t),\quad \dot{\theta}=\sqrt{2D_R}\,\eta (t),
\end{align}
where $v_0$ is the self-propulsion speed and $D_R$ is the rotational diffusion coefficient.
We consider the initial condition where the particle starts at the origin with a random orientation $\theta(0)$ chosen uniformly from $[-\pi,\pi]$. As a result the position distribution is isotropic at all times, and it is enough to consider the $(x,\theta)$ process only. The Fokker-Planck equation governing the joint probability distribution $P(x,\theta,t)$ is given by,
\begin{align}
\frac{\partial P}{\partial t}=-v_0\cos\theta\frac{\partial P}{\partial x}+D_R\frac{\partial ^2 P}{\partial \theta^2},
\label{abp:FP}
\end{align}
with the initial condition $P(x,\theta,0)=\delta(x)/(2\pi).$ 

Similar to the other active processes discussed above, at times much larger than the characteristic time $D_R^{-1}$, the process $x(t)$ becomes diffusive with an effective diffusion constant $D_\ab=v_0^2/(2D_R)$~\cite{Basu2019}. Therefore, anticipating the diffusive behavior at long times, we rewrite \eref{abp:FP} in terms of the scaled variable $y= x/\sqrt{D_\ab}$.
\begin{align}
\varepsilon^2 \frac{\partial P}{\partial t}=-\varepsilon \sqrt{2} \cos\theta \frac{\partial P}{\partial y}+\mathcal{L}_\theta P,\text{ with } \mathcal{L}_\theta=\frac{\partial ^2}{\partial \theta^2},
\label{e:abp}
\end{align}
where $\varepsilon^2=1/D_R$ is the persistence time of the $\theta$ dynamics.

The isotropic  initial condition for the orientation ensures that the marginal position distribution $\rho(y,t)=\int_{-\pi}^{\pi}d\theta P(y,\theta,t)$ is symmetric at all times, i.e., distribution $\rho(-y,t)=\rho(y,t)$. This implies that all the odd moments of the position vanish. Moreover, \eref{e:abp} remains invariant under the transformation $(y,\varepsilon)\to (-y,-\varepsilon)$, as a result, $\rho(y,t)$ contains only even powers of $\varepsilon$.

As mentioned in Sec.~\ref{Sec:genstrategy}, we need the $2n$-th moment $\la y^{2n}(t)\ra$ to determine the distribution uniquely at the order $(\varepsilon^2/t)^n$; so in the following, we first discuss the computation of the moments recursively (see also \cite{Shee2022_ddim}).

\subsection{Moments}\label{Sec.abp:moments}
To compute the moments $\la y^{2n}(t)\ra$, it is useful to define the correlation functions,
\begin{align}
M(k,n,t)=\la y^k \cos(n\theta) \ra=\int_{-\infty}^{\infty}dy\,\int_{-\pi}^{\pi}d\theta \, y^k \cos(n\theta) P(y,\theta,t).
\end{align}
Note that, $M(k,0,t)=\la y^{k}(t)\ra$ are the $k$th position moments. Multiplying both sides of the  \eref{e:abp} by $y^{k}
\cos(n\theta)$ and integrating over $y$ and $\theta$, we get, for $n,k\geq 1$,
\begin{align}
\left[\varepsilon^2 \frac{d}{dt}+n^2\right]M(k,n,t)=\varepsilon\frac{k}{\sqrt{2}}\left( M(k-1,n-1,t)+M(k-1,n+1,t)  \right).
\label{abp_momentskn}
\end{align}
For $n=0$ and $k\geq 1$,
\begin{align}
\varepsilon^2 \frac{d}{dt}M(k,0,t)=\sqrt{2}\,k\varepsilon\, M(k-1,1,t).
\label{abp_momentsk0}
\end{align}
The initial conditions for Eqs.~\eqref{abp_momentskn} and \eqref{abp_momentsk0} is $M(k,n,0)=0.$ Moreover, the normalization condition and $\la\cos(n\theta)\ra=0$ leads to,
\begin{align}
M(0,0,t)=1\quad\text{and}\quad M(0,n,t)=0 \text{ for } n>0\text{  respectively}.
\label{abp_m0nt}
\end{align}

Using the initial conditions, the solutions for $n\geq 1$ and $k\geq 1$ are,
\begin{align}
M(k,n,t)&=\frac{k}{\sqrt{2}\,\varepsilon}\int_0^t dt'e^{-\frac{(t-t')n^2}{\varepsilon^2}} \left[ M(k-1,n-1,t')+M(k-1,n+1,t')  \right].
\label{abp:momentsInt}
\end{align}
The solution of \eref{abp_momentsk0} yields for the positions moments,
\begin{align}
M(k,0,t)=\frac{k\sqrt{2}}{\varepsilon}\int_0^t dt' M(k-1,1,t').
\end{align}
\begin{figure}
\centering\includegraphics[width=0.7\hsize]{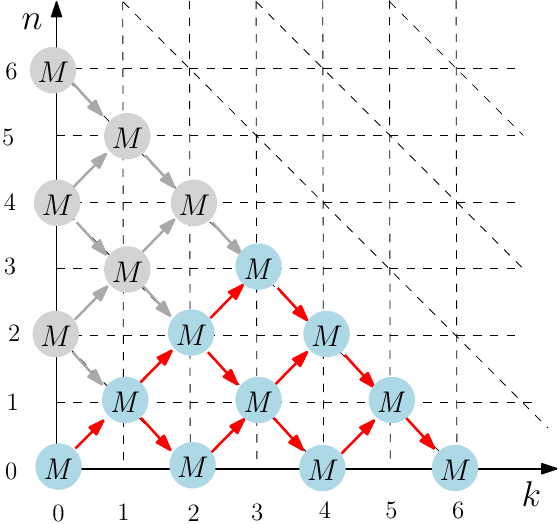}
\caption{Illustration of the recursive connections between the correlation functions $M(k,n,t)$ for different $k,\,n$ on the $(k,n)$ plane with even $k+n$; see \eref{abp:momentsInt}.}
\label{f:abpmoments}
\end{figure}
It follows from the structure of \eref{abp:momentsInt} that the recursive connections between the correlation functions $\{M(k,n,t)\}$ for different values of $(k,n)$ form two independent networks, which sit on the even and odd $k+n$ sub-lattice respectively. Since on the $k=0$ boundary line, $M(0,0,t)=1$ is the only nonzero term, it follows that $M(k,n,t)$ is zero for all odd $k+n$. Therefore, to determine the non-zero position moments $M(k,0,t)$ for even $k$, we need to consider the even $k+n$ network only.
This network is illustrated in \fref{f:abpmoments} with the relevant recursive connections. From the figure it is further clear that the correlation functions $M(k,n,t)$ vanish for $n>k$. On the line $n=k$, the correlations simplify to,
\begin{equation}
M(k,k,t)=\frac{k}{\sqrt{2}\,\varepsilon}\int_0^t dt' e^{-\frac{(t-t')k^2}{\varepsilon^2}}M(k-1,k-1,t').
\label{abp:mkk}
\end{equation}

 Let us illustrate the recursive procedure by computing the position variance $M(2,0,t)$ explicitly. To this end, we first need $M(1,1,t)$ (see \fref{f:abpmoments}), which is straightforwardly obtained from  \eref{abp:mkk} as,
\begin{equation}
M(1,1,t)=\frac{\varepsilon }{\sqrt{2}}\left(1-e^{-\frac{t}{\varepsilon ^2}}\right).
\end{equation}
Using this, from \eref{drabp:positionmoments}, we get,
\begin{equation}
M(2,0,t)=2t -2\varepsilon ^2 \left(1-e^{-\frac{t}{\varepsilon ^2}}\right).\label{abp:m20t}
\end{equation}

For $M(4,0,t)$, we now need to compute two additional correlation functions $M(2,2,t)$ and $M(3,1,t)$, as the remaining required correlation functions have already been computed during the previous evaluation of $M(2,0,t)$. In general, computation of the moment $M(2j-2,0,t)$, requires all the correlation functions in the triangular region $k\geq n$, $k+n\leq 2j-2$ and $n\geq 0$, on the even $k+n$ sub-lattice. Therefore, subsequent computation of $M(2j,0,t)$ requires the computation of only $j$ additional correlation functions along the $k+n=2j$ line, starting with $k=n=j$.
Following this procedure, we find the fourth and sixth position moments as,
\begin{align}
M(4,0,t)&=12 t^2-5\varepsilon^2t\left(9+4e^{-\frac{t}{\varepsilon^2}}\right)+\frac{\varepsilon^4}{12}\left(783-784e^{-\frac{4t}{\varepsilon^2}}+ e^{-\frac{4t}{\varepsilon^2}}\right),
\label{abp:m4}
\end{align}
%\begin{align}
%M(4,0,t)&=12 t^2-\varepsilon^2\left(45t+\frac{2(98\varepsilon^2+90t)}{3}e^{-\frac{t}{\varepsilon^2}}\right)+\frac{\varepsilon^4}{4}\left(261+\frac{1}{3}e^{-\frac{4t}{\varepsilon^2}}\right),
%\label{abp:m4}
%\end{align}
%\begin{align}
%M(4,0,t)&=12 t^2-45 \varepsilon ^2 t+\frac{261 \varepsilon ^4}{4}
%-\frac{4\varepsilon ^2}{3}  \left(49 \varepsilon ^2+15 t\right) e^{-\frac{t}{\varepsilon ^2}}+\frac{\varepsilon^4}{12}e^{-\frac{4 t}{\varepsilon ^2}},
%\label{abp:m4}
%\end{align}
and
\begin{align}
M(6,0,t)&=120 t^3-10\varepsilon^2t^2\left(99-25e^{-\frac{t}{\varepsilon ^2}} \right)+\varepsilon^4 t\left( \frac{7495}{2}+\frac{13615}{6} e^{-\frac{t}{\varepsilon^2}}-\frac{1}{6}e^{-\frac{4t}{\varepsilon^2}}\right)\cr
&+\frac{\varepsilon^6}{6}\left(\frac{108295}{3}+\frac{288785}{8}  e^{-\frac{t}{\varepsilon^2}}+\frac{1}{5}e^{-\frac{4t}{\varepsilon^2}}+\frac{1}{120}e^{-\frac{9t}{\varepsilon^2}} \right),
\label{abp:m6}
\end{align}
respectively. We will need the position moments calculated above to determine the position distribution perturbatively in $\varepsilon$.

\subsection{Position distribution}
We now proceed to compute the long-time position distribution perturbatively. For this purpose, it is important to first note that, the distribution of $\theta$, evolving by the Fokker-Planck operator $\mathcal{L}_\theta$, reaches a steady state. Thus, we can always express the solution  %\eref{e:underdampedfp}  
of \eref{e:abp}  the eigenbasis of $\mathcal{L}_\theta$ as,
 \bea
 P(y,\theta,t)=\sum_{n=0}^{\infty}p_n(\theta) F_n(y,t),
 \label{cosbasis}
 \eea
where,
\bea
p_0(\theta)=\frac{1}{2\pi}\quad \text{and}\quad p_n(\theta)=\frac{1}{\pi}\cos(n\theta) \text{ for }n\geq 1
\label{abp:pntheta}
\eea
are the eigenfunctions of $\mathcal{L}_\theta$ with eigenvalue $-n^2$,
\bea
\mathcal{L}_\theta p_n(\theta)=-n^2p_n(\theta).
\eea
They obey the following orthonormality relations.
\begin{align}
\int_{-\pi}^{\pi}d\theta\, \cos(m\theta)\,p_n(\theta)=&\delta_{m,n},\\
\int_{-\pi}^{\pi}d\theta\cos\theta\,\cos(m\theta)\,p_n(\theta)=&\frac{1}{2}(\delta_{n-1,m}+\delta_{n+1,m}).
\label{abp:pntheta_ortho}
\end{align}
Note that, integration of \eref{cosbasis} with respect to $y$ yields,
\bea
\int_{-\infty}^{\infty} P(y,\theta,t)\,dy=p_0(\theta),
\label{abp:pthetat}
\eea
since we start with the stationary initial condition on $\theta$. In general, if we began from some particular $\theta$, then,
\bea
\int_{-\infty}^{\infty} P(y,\theta,t)\,dy=\sum_{n=0}^{\infty}p_n(\theta)\, e^{-n^2 t/\varepsilon^2}.
\eea

Our goal is to obtain the marginal position distribution,
\bea
 \rho(y,t)=\int_{-\infty}^{\infty}P(y,\theta,t)\,d\theta= F_0(y,t).
 \eea
 In general, $F_n(y,t)$ is given by,
 \bea
 F_n(y,t)=\int_{-\infty}^{\infty}P(y,\theta,t)\cos(n\theta)\,d\theta.
 \eea
 However, it must be remembered that the above equation cannot be used as the joint distribution $P(y,\theta,t)$ is unknown.

Substituting \eref{cosbasis} in \eref{e:abp}, we get,
\begin{align}
\varepsilon^2\sum_{n=0}^{\infty} p_n(\theta)\frac{\partial F_n}{\partial t}=-\varepsilon \sqrt{2}\sum_{n=0}^{\infty} \cos\theta\, p_n(\theta)\frac{\partial F_n}{\partial y}-\sum_{n=0}^{\infty} n^2 p_n(\theta) F_n.
\label{abp:delta1}
\end{align}
Multiplying \eref{abp:delta1} on both sides by $\cos(m\theta)$ and integrating w.r.t. $\theta$ thereafter, we get,
\begin{subequations}
\label{abp:ams}
\begin{align}
\varepsilon^2 \frac{\partial F_0}{\partial t}&=-\varepsilon\sqrt{2}\,\frac{\partial F_1}{\partial y}\quad\text{for }m=0,
\label{abp:a0}\\
\left[\varepsilon^2 \frac{\partial }{\partial t}+m^2\right]F_m&=-\frac{\varepsilon}{\sqrt{2}}\left( \frac{\partial F_{m+1}}{\partial y}+\frac{\partial F_{m-1}}{\partial y} \right)\quad\text{for }m>0.
\label{abp:am}
\end{align}
\end{subequations}
From \eref{abp:pthetat}, it is clear that $F_m(y,t)$ does not have a series in $\nu$, unlike the AOUP case. 
So, we look directly for series solution of $F_{m}(y,t)$ in the form,
\begin{align}
F_{m}(y,t)=\sum_{k=0}^{\infty}\varepsilon^k A_m^k(y,t).
\label{abp:amk}
\end{align}

Putting \eref{abp:amk} in Eqs.~\eqref{abp:a0} and \eqref{abp:am} and collecting terms of the order $\varepsilon^k$, we have,
\begin{align}
\frac{\partial A_0^{k-2}}{\partial t}=&-\sqrt{2}\frac{\partial A_1^{k-1}}{\partial y},\label{abprec1}\\
\frac{\partial A_m^{k-2}}{\partial t}=&-\frac{1}{\sqrt{2}}\left(  \frac{\partial A_{m+1}^{k-1}}{\partial y}+\frac{\partial A_{m-1}^{k-1}}{\partial y} \right)-m^2A_m^k,\quad m>0,
\label{abprec2}
\end{align}
with $A_m^k(y,t)=0$ for $k<m$. Before finding the solutions of $A_m^k$s, it is useful to simplify the series in \eref{abp:amk}---in particular, in the following we show that $A_m^k(y,t)$ is non-zero only for even $m+k$.

First, we note that $F_{0}(y,t)$, which is the marginal distribution $\rho(y,t)$ at large times, is an even function of $y$ at all times due to the symmetric initial conditions. Again \eref{abp:ams} is symmetric under $(y,\varepsilon)\to(-y,-\varepsilon)$. Thus $F_{0}$ contains only even powers of $\varepsilon$, i.e., 
\begin{align}
\rho(y,t)=F_{0}(y,t)=\sum_{k=0}^{\infty}\varepsilon^{2k}A_0^{2k}(y,t).
\label{abp:marginal}
\end{align}
Now, putting $k=0$, in \eref{abprec2}, we have, $m^2 A_m^0=0$. Thus,
\begin{align}
A_m^0=\delta_{m,0}A_0^0.
\label{abp:a00}
\end{align}
\begin{figure}
\centering\includegraphics[width=0.7\hsize]{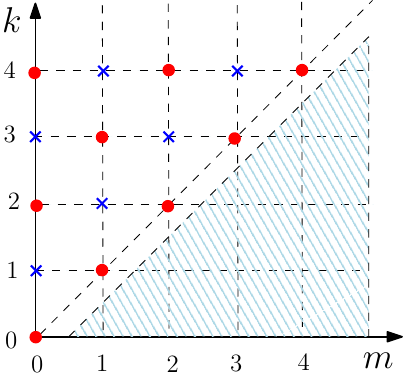}
\caption{Graphical representation of recursive determination of  $A_m^k$, given by \eref{abprec1} and \eref{abprec2}. The red dots represent the non-zero $A_m^k$, whereas the blue cross represents $A_m^k=0$.}
\label{f:abp_am}
\end{figure}

Next, putting $k=1$ in \eref{abprec2},
\begin{align}
m^2 A_m^1=-\frac{1}{\sqrt{2}}\left(  \frac{\partial A_{m+1}^{0}}{\partial y}+\frac{\partial A_{m-1}^{0}}{\partial y} \right).
\end{align}
The above equation combined with \eref{abp:a00} leads to,
\begin{align}
A_m^1=-\frac{1}{\sqrt{2}}\delta_{m,1}\frac{\partial A_{0}^{0}}{\partial y}.
\end{align}
In fact, one can systematically proceed by putting $k=2,3,4,\dots$ and find the non-zero $A_m^k$ in the $(m,k)$ plane. This process is best illustrated in graphically in \fref{f:abp_am}. The fact that $A_0^0(y,t)\neq 0$ and $A_0^1(y,t)=0$, recursively leads to $A_m^k(y,t)=0$ for odd $m+k$ and $m>k$. Thus, \eref{abp:amk} can be refined as,
\begin{align}
F_{m}(y,t)=\sum_{k=m}^{\infty}\varepsilon^k A_m^k(y,t)\quad\text{with }(m+k)\text{ even.}
\label{abp:amk2}
\end{align}

We proceed to obtain $A_0^{2k}(y,t)$, for different $k$, which in general satisfies the differential equation, 
\begin{align}
\frac{\partial A_0^{2k}}{\partial t}=&-\sqrt{2}\frac{\partial A_1^{2k+1}}{\partial y}.
\label{abp:a02k}
\end{align}
For $k=0$, we have,
\begin{align}
\frac{\partial A_{0}^{0}}{\partial t}=-\sqrt{2}\frac{\partial A_{1}^{1}}{\partial y}.
\label{abp:a00_1}
\end{align}
In general, we find, using \eref{abprec2}, $A_m^m(y,t)$ is related to $A_0^0(y,t)$ via a simple relation,
\begin{align}
A_m^m=&-\frac{1}{m^2\sqrt{2}}\frac{\partial A_{m-1}^{m-1}}{\partial y}=\frac{(-1)^m}{2^{m/2}(m!)^2}\frac{\partial^m A_{0}^{0}}{\partial y^m}.
\label{abp:amm}
\end{align}
Substituting $A_{1}^{1}(y,t)$ from \eref{abp:amm} in \eref{abp:a00_1}, yields the diffusion equation for $A_0^0(y,t)$,
\begin{align}
\frac{\partial A_{0}^{0}}{\partial t}=\frac{\partial^2 A_{0}^{0}}{\partial y^2}.
\end{align}
Thus we get the normalized marginal position distribution to order $\varepsilon^0$ as,
\begin{align}
A_0^0(y,t)=\frac{1}{\sqrt{4\pi t}}\exp\left(-\frac{y^2}{4t}\right).\label{abp:a00_sol}
\end{align}
For the next order correction to the marginal distribution $A_0^2(y,t)$, we get from \eref{abp:a02k},
\begin{align}
\frac{\partial A_{0}^{2}}{\partial t}=-\sqrt{2}\frac{\partial A_{1}^{3}}{\partial y}.
\end{align}
To obtain a closed differential equation for $A_0^2(y,t)$, we need to express the right hand side in terms of $A_0^2(y,t)$ and  already known functions namely $A_0^0(y,t)$. To this end, we put $m=1$, $k=3$ in \eref{abprec2}, to obtain,
\begin{align}
A_1^3=-\frac{1}{\sqrt{2}}\left(\frac{\partial A_0^2}{\partial y}+\frac{\partial A_2^2}{\partial y}+\sqrt{2}\frac{\partial A_1^1}{\partial t}\right).
\end{align}
Now, we can again use the recursion relation \eref{abp:amm} to express $A_1^1(y,t)$ and $A_2^2(y,t)$ recursively in terms of $A_0^0(y,t)$, to get an inhomogeneous diffusion equation [similar to \eqref{strategy:inhomo_diff}] for $A_0^2(y,t)$,
\begin{align}
\Bigg[\frac{\partial }{\partial t}-\frac{\partial^2 }{\partial y^2}\Bigg]A_{0}^{2}=-\frac{\partial^2}{\partial y^2}\left(\frac{\partial A_{0}^{0}}{\partial t}-\frac{1}{8}\frac{\partial^2 A_{0}^{0}}{\partial y^2} \right),
\end{align}
where the source terms on the right hand side depend on the $\varepsilon^0$ order solution $A_0^0(y,t)$. To solve this inhomogeneous diffusion equation, we anticipate a solution of the form,
\begin{align}
A_0^{2}(y,t)=\frac{1}{t}\,q_{2}\left(\frac{y}{\sqrt{4t}}\right)\frac{e^{-y^2/(4t)}}{\sqrt{4\pi t}},\label{abp:a02_scaling}
\end{align}
owing to the diffusive scaling at the leading order $\varepsilon^0$. Infact, in general, for higher orders,
\begin{align}
A_0^{2k}(y,t)=\frac{1}{t^{k}}\,q_{2k}\left(\frac{y}{\sqrt{4t}}\right)\frac{e^{-y^2/(4t)}}{\sqrt{4\pi t}}.\label{abp:a02k_scaling}
\end{align}
Using \eref{abp:a00_sol} and scaling form \eref{abp:a02_scaling}, we get an inhomogeneous Hermite differential equation [as mentioned in \eref{gen:hermite_inh}] for $q_2(z)$,
\begin{align}
q_2''(z)-2zq_2'(z)+4q_2(z)=\frac{7}{8}(4z^4-12z^2+3).
\end{align}
The above inhomogeneous Hermite equation can be solved easily, using \eref{eq:qn_gen} to get $q_2(z)$,
\begin{align}
q_2(z)=C_2\,H_{2}(z)+\frac{1}{8}\left(\frac{21}{2}z^2-7z^4\right).\label{abp:q2_gen}
\end{align}
The normalization condition $\int_{-\infty}^{\infty}dy A_0(y,t)$ is trivially satisfied for arbitrary values of $C_2$ and thus, as mentioned before, we take recourse to the moments to evaluate $C_2$. In fact, at each order, the constant $C_{2k}$ is determined by comparing the coefficient of $(\tau/t)^k$ of $M(2k,0,t)/(4t)^k$ obtained from the two methods: the exact computation in Sec.~\ref{Sec.abp:moments} and using the series \eref{abp:marginal}, where the latter is simply given by,
\begin{align}
\int_{-\infty}^{\infty}dz\, z^{2k} q_{2k}(z)e^{-z^2}/\sqrt{\pi}.
\label{abp:m2k_series}
\end{align}
Following this procedure for $k=1$, we get $C_2=5/64$, which leads to,
\begin{align}
q_2(z)=-\frac{1}{32}\left(5-52z^2+28z^4\right).
\label{abp:q2F}
\end{align}

For the next subleading contribution $A_0^4(y,t)$ to the marginal distribution we can again start from \eref{abp:a02k} and use \eref{abprec2} to obtain,
\begin{align}
\Bigg[\frac{\partial }{\partial t}-\frac{\partial^2 }{\partial y^2}\Bigg]A_0^4(y,t)=S_4(y,t),
\end{align}
where the inhomogeneous term $S_4(y,t)$ depends on the solution of previous orders.
\begin{align}
S_4(y,t)=\frac{\partial^2 }{\partial y^2}\left(-\frac{\partial }{\partial t}   +\frac{1}{8}\frac{\partial^2 }{\partial y^2}\right)A_0^2+\frac{\partial^2 }{\partial y^2}\left(\frac{\partial^2}{\partial t^2} -\frac{9}{32}\frac{\partial^3}{\partial y^2\partial t}+\frac{5}{288}\frac{\partial^4}{\partial y^4} \right)A_0^0.
\end{align}

Considering $A_0^4(y,t)$ to be of the form \eref{abp:a02k_scaling}, we get an inhomogeneous equation for $q_4(z)$,
\begin{align}
q_4''(z)-2zq_4'(z)+8q_4(z)=s_4(z),
\end{align}
where the inhomogeneous term $s_4(z)=-\frac{1115}{768}+\frac{1565}{32}z^2-\frac{8275}{96}z^4+\frac{2371}{72}z^6-\frac{49}{16}z^8$. The solution of the above equation can be obtained using \eref{eq:qn_gen},
\begin{align}
q_4(z)=C_4 H_4(z)+\frac{49 z^8}{128}-\frac{1655 z^6}{576}+\frac{19895 z^4}{4608}-\frac{1115 z^2}{1536}.
\end{align}
The constant $C_4$, obtained by comparing the coefficient of $(\tau/t)^2$ in the expansion of $M(4,0,t)/(4t)^2$ in \eref{abp:m4} with \eref{abp:m2k_series} (for $k=2$) in the $t\gg\tau$ limit, turns out to be $-677/73728$. This yields,
\begin{align}
q_4(z)=-\frac{677}{6144}-\frac{73}{256}z^2+\frac{3203}{768}z^4-\frac{1655}{576}z^6+\frac{49}{128} z^8.
\label{abp:q4F}
\end{align}

Proceeding in a similar way, we find the subsequent sub-leading contribution $A_0^6(y,t)$ satisfy the inhomogeneous differential equation,
\begin{align}
\Bigg[\frac{\partial }{\partial t}-\frac{\partial^2 }{\partial y^2}\Bigg]A_0^6(y,t)=S_6(y,t),\label{abp:a06eq}
\end{align}
where the inhomogeneous term $S_6(y,t)$ depends on the previous order solutions $A_0^{0,2,4}(y,t)$. Substituting the scaling ansatz \eref{abp:a02k_scaling}, we get an inhomogeneous equation for $q_6(z)$, whose general solution is given by \eqref{eq:qn_gen} in terms of an undetermined constant $C_6$. This undetermined constant can again be obtained by comparing the moment $M(6,0,t)$ obtained from the two ways, as in the previous orders. Skipping details (see~\ref{Sec:ABPa06}), we get,
\begin{align}
q_6(z)=-&\frac{302975}{1769472}-\frac{92375 }{147456}z^2+\frac{293635 }{442368}z^4+\frac{1790095}{165888} z^6-\frac{325727}{36864} z^8\cr
&+\frac{17437 }{9216}z^{10}-\frac{343 }{3072}z^{12}.
\label{abp:q6final}
\end{align} 
We can go on and calculate the higher order contributions following the same procedure as described above. 

As we have mentioned in the introduction, the exact position distribution of ABP is known as an infinite series in Fourier space in terms of the eigenvalues and eigenfunctions of the Mathieu equations\cite{Franosch2016,Basu2019}. In \ref{mathieu} we extract the subleading contributions to the real-space position distribution from this infinite series and show that they match with the results obtained in this section.

\section{Direction reversing active Brownian particle}\label{Sec:DRABP}
Direction reversing active Brownian particles (DRABP) models the motion of a certain class of bacteria like {\it Myxococcus xanthus} and {\it Pseudomonas putida}. The stochastic evolution of the position $(x,y)$ of a DRABP in two-dimensions is governed by,
\begin{align}
\dot{x}(t)&= v_0\, \sigma(t) \cos \theta(t),\quad \dot{y}(t)= v_0 \,\sigma(t)\sin \theta(t), \quad\dot{\theta}(t)=\sqrt{2D_R}~\eta (t),
\end{align}
where the dichotomous $\sigma$ alternates between $\pm 1$ at a rate $\gamma$, while the internal orientation vector $\theta$ undergoes a rotational diffusion with diffusion constant  $D_R$. We consider the initial condition where the particle starts at the origin with a random orientation $\theta(0)$ chosen uniformly from $[-\pi,\pi]$ and $\sigma(0)=\pm 1$ with equal probability $1/2$. As a result the position distribution is isotropic at all times, and it is enough to look at the $x$-position only. So for simplicity, we consider only the $(x,\sigma,\theta)$ process, which is also a Markov process. The corresponding Fokker-Planck equation for $P_\sigma(x,\theta,t)$ is given by,
\begin{align}
\frac{\partial P_\sigma}{\partial t}=-v_0\sigma\cos\theta\frac{\partial P_\sigma}{\partial x}+D_R\frac{\partial ^2 P_\sigma}{\partial \theta^2}-\gamma P_{\sigma}+\gamma P_{-\sigma},
\label{drabp:fp1}
\end{align}
with the initial condition,
\begin{equation}
P_{\sigma}(x,\theta,0)=\frac{1}{2\pi}\,\delta(x)\,\left[\frac{\delta_{\sigma,1}+\,\delta_{\sigma,-1}}{2}\right].
\label{drabp_IC}
\end{equation}
It is convenient to write \eref{drabp:fp1} as,
\begin{align}
\frac{\partial P}{\partial t}&=-v_0\cos\theta\frac{\partial Q}{\partial x}+D_R\frac{\partial ^2 P}{\partial \theta^2},\\
\frac{\partial Q}{\partial t}&=-v_0\cos\theta\frac{\partial P}{\partial x}+D_R\frac{\partial ^2 Q}{\partial \theta^2}-2\gamma Q,
\end{align}
where $P=P_+ +P_-$ and $Q=P_+-P_-$.

Using the effective noise correlation, we earlier argued that~\cite{Santra2021}, the process $x(t)$, at times much longer than the correlation-time, becomes diffusive with the effective diffusion coefficient $D_{\dr}=v_0^2/[2(D_R+2\gamma)].$ Therefore, anticipating the diffusive scaling at long times, we rewrite the above equations in terms of the scaled variable $y=x/\sqrt{D_\dr}$, as,
\begin{subequations}
\label{fp:drabp_delta}
\begin{align}
\varepsilon^2 \frac{\partial P}{\partial t}&=-\varepsilon \sqrt{2(\lambda+1)} \cos\theta \frac{\partial Q}{\partial y}+\mathcal{L}_\theta P,\\
\varepsilon^2 \frac{\partial Q}{\partial t}&=-\varepsilon \sqrt{2(\lambda+1)} \cos\theta \frac{\partial P}{\partial y}+\mathcal{L}_\theta Q-\lambda Q,
\end{align}
\end{subequations}
where the operator $\mathcal{L}_\theta=\partial ^2/\partial \theta^2$, $\varepsilon^2=1/D_R$ is the persistence-time of the $\theta$ dynamics,and the dimensionless parameter $\lambda=2\gamma/D_R$ denotes the ratio of the rotational diffusion and directional reversal time-scales.

 The isotropic initial condition \eref{drabp_IC} leads to a symmetric marginal distribution  $\rho(y,t)\equiv\int_{-\pi}^{\pi} d\theta P(y,\theta,t)=\rho(-y,t)$. As a result all the odd moments  of position vanish.  Moreover, it follows from \eref{fp:drabp_delta} that $P(y,\theta,t)$ and $Q(y,\theta,t)$ remain invariant under the transformation $(y,\varepsilon)\to(-y,-\varepsilon)$. Consequently, $\rho(y,t)$ must contain only even powers $\varepsilon$.
 
 As we have mentioned earlier, the $2n$-th moment $\la y^{2n}(t)\ra$ is needed to completely determine the distribution at order $(\varepsilon^2/t)^n$; we first determine the moments recursively in the next section.

\subsection{Moments}\label{sec:drabpmoments}
To determine the position moments $\la y^{2n}(t)\ra$, it is convenient to define the following \emph{correlation functions},
\begin{subequations}
\begin{align}
M(k,n,t)=\int_{-\infty}^{\infty}dy\,\int_{-\pi}^{\pi} d\theta\, y^k\, \cos(n\theta)\, P(y,\theta,t),\\
L(k,n,t)=\int_{-\infty}^{\infty}dy\,\int_{-\pi}^{\pi} d\theta\, y^k\, \cos(n\theta)\, Q(y,\theta,t),
\end{align}
\end{subequations}
such that $M(k,0,t)=\la y^{k}(t)\ra$. Here, both $n$ and $k$ are non-negative integers.

Multiplying both sides of Eqs.~\eqref{fp:drabp_delta} by $y^k\cos(n\theta)$ and then integrating over $y$ and $\theta$, we get, for $n,k\geq 1$
\begin{subequations}
\label{drabp_moment_kn}
\begin{align}
\left[\varepsilon^2 \frac{d}{dt}+n^2\right]M(k,n,t)&=k\varepsilon\sqrt{\frac{\lambda+1}{2}}\,\Big[ L(k-1,n-1,t)+L(k-1,n+1,t)  \Big],\\
\left[\varepsilon^2 \frac{d}{dt}+n^2+\lambda\right] L(k,n,t)&=k\varepsilon\sqrt{\frac{\lambda+1}{2}}\,\Big[ M(k-1,n-1,t)+M(k-1,n+1,t)  \Big].
\end{align}
\end{subequations}
For $n=0$ and $k\geq 1$, we have,
\begin{subequations}
\label{drabp_moment_n0}
\begin{align}
\varepsilon^2 \frac{d}{dt}M(k,0,t)&=k\varepsilon\sqrt{2(\lambda+1)}\,L(k-1,1,t),\label{drabp_moment_n0A}\\
\left[\varepsilon^2 \frac{d}{dt}+\lambda\right] L(k,0,t)&=k\varepsilon\sqrt{2(\lambda+1)}\,M(k-1,1,t) .
\end{align}
\end{subequations}
The initial conditions for Eqs.~\eqref{drabp_moment_kn}-\eqref{drabp_moment_n0} are $M(k,n,0)=L(k,n,0)=0$. Moreover,
from the normalization and $\la\cos(n\theta)\ra=0$, it respectively follows that,
\begin{align} 
 M(0,0,t)=1\quad\text{and}\quad M(0,n,t)=L(0,n,t)=0 \text{ for } n>0,
 \label{drabp:moments_ic}
\end{align} 
for all $t$.
 Using the initial conditions, the solutions for $n\geq 1$ and $k\geq 1$ are,
\begin{subequations}
\label{drabp:momentssol}
\begin{align}
M(k,n,t)&=\frac{k}{\varepsilon}\sqrt{\frac{\lambda+1}{2}}\int_0^t dt'e^{-\frac{(t-t')n^2}{\varepsilon^2}} \left[ L(k-1,n-1,t')+L(k-1,n+1,t')  \right]\\
L(k,n,t)&=\frac{k}{\varepsilon}\sqrt{\frac{\lambda+1}{2}}\int_0^t dt'e^{-\frac{(t-t')(n^2+\lambda)}{\varepsilon^2}} \left[ M(k-1,n-1,t')+M(k-1,n+1,t')  \right].
\end{align}
\end{subequations}

From \eref{drabp_moment_n0A}, the solution for the position moments $M(k,0,t)$ can be written as,
 \begin{equation}
M(k,0,t)=\frac{k}{\varepsilon}\sqrt{2(\lambda+1)}\int_0^t dt'  L(k-1,1,t').
\label{drabp:positionmoments}
 \end{equation}

%%%%%%%%%%%%%%%%%%%%%%%%%%% 
 \begin{figure}
 \centering\includegraphics[width=0.7\hsize]{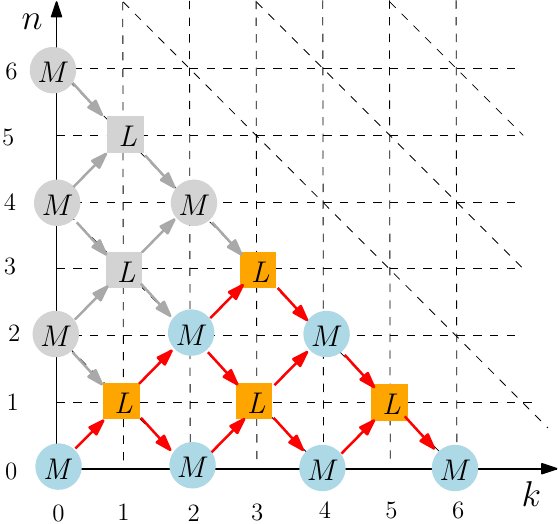}
 \caption{The illustration of recursive connections between the correlation functions $\{M(k,n,t),L(k,n,t)\}$, for different values of $(k,n)$ with even $k+n$; see \eref{drabp:momentssol}. In the $n>k$ sector, shown in grey, $M(k,n,t)=L(k,n,t)=0$.}
 \label{drabp:MLconF}
 \end{figure}
%%%%%%%%%%%%%%%%%%%%%%%%%%% 

The integral equations \eref{drabp:momentssol} can be used recursively to obtain the position moments $M(k,0,t)$ from \eref{drabp:positionmoments}. It is evident from the structure of these equations, that recursive connections  between the correlation functions $\{M(k,n,t),L(k,n,t)\}$, for different values of $(k,n)$, form two independent networks, sitting on even and odd $k+n$ sub-lattices.
 Since, on the $k=0$ boundary, the only non-zero term is $M(0,0,t)=1$, it follows that $M(k,n,t)=L(k,n,t)=0$ when $k+n$ is odd. Therefore, to determine the non-zero position moments $M(k,0,t)$ for even $k$, we need to stay on the even $k+n$ network. This network, with the relevant recursive connections is illustrated in Fig.~\ref{drabp:MLconF}. From this figure, it is further clear that $M(k,n,t)$ and $L(k,n,t)$ also vanish for $n>k$. On the $n=k$ boundary, \eref{drabp:momentssol} simplifies to,
\begin{subequations}
\label{drabp:mkk}
\begin{align}
M(k,k,t)&=\frac{k}{\varepsilon}\sqrt{\frac{\lambda+1}{2}}\int_0^t dt'e^{-\frac{(t-t')k^2}{\varepsilon^2}}  L(k-1,k-1,t')\\
L(k,k,t)&=\frac{k}{\varepsilon}\sqrt{\frac{\lambda+1}{2}}\int_0^t dt'e^{-\frac{(t-t')(k^2+\lambda)}{\varepsilon^2}}  M(k-1,k-1,t').
\end{align}
\end{subequations}
We illustrate the recursive procedure by computing the position variance $M(2,0,t)$ explicitly. To this end, we first need $L(1,1,t)$ (see \fref{drabp:MLconF}), which is straightforwardly obtained from  \eref{drabp:mkk} as,
\begin{equation}
L(1,1,t)=\frac{\varepsilon }{\sqrt{2(\lambda+1)}}\left(1-e^{-\frac{(\lambda+1) t}{\varepsilon ^2}}\right).
\end{equation}
Using this, from \eref{drabp:positionmoments}, we get,
\begin{equation}
M(2,0,t)=2t -\frac{2\varepsilon ^2 }{\lambda+1}\left(1-e^{-\frac{(\lambda+1) t}{\varepsilon ^2}}\right).
\end{equation}

For $M(4,0,t)$, we now need to compute two additional correlation functions $M(2,2,t)$ and $L(3,1,t)$, as the remaining required correlation functions have already been computed during the previous evaluation of $M(2,0,t)$. In general, computation of the moment $M(2j-2,0,t)$, requires all the correlation functions in the triangular region $k\geq n$, $k+n\leq 2j-2$ and $n\geq 0$, on the even $k+n$ sub-lattice. Therefore, subsequent computation of $M(2j,0,t)$ requires the computation of only $j$ additional correlation functions along the $k+n=2j$ line, starting with $k=n=j$.
Following this procedure, we also evaluate the fourth and sixth position moments as,
\begin{align}
M(4,0,t)=&12t^2-\frac{3\varepsilon^2t}{4(\lambda+1)}\left[4(\lambda-15)+e^{-\frac{(\lambda+1)t}{\varepsilon^2}}\frac{16(3\lambda-5)}{(3-\lambda)}  \right]\cr
+&\frac{3\varepsilon^4}{4(\lambda+1)^2}\Bigr[87-10\lambda -\lambda^2-16e^{-\frac{(\lambda+1)t}{\varepsilon^2}}\frac{(49-38\lambda+9\lambda^2)}{(\lambda-3)^2}+e^{-\frac{4t}{\varepsilon^2}}(\lambda+1)^4  \Bigr],\cr &\label{drabp:m4t}
\end{align}
and
\begin{align}
M(6,0,t)=&120t^3+\frac{90 \varepsilon^2t^2}{\lambda+1}\left[\lambda-11+e^{-\frac{(\lambda+1)t}{\varepsilon^2}}\frac{(3\lambda-5)^2}{(\lambda-3)^2}\right]
+\frac{45\varepsilon^4t}{(\lambda+1)^2}\cr
\times & \left[ \frac{67455 - 3375 \lambda - 1755 \lambda^2 - 45 \lambda^3 }{2  (\lambda+9)} +e^{-\frac{4t}{\varepsilon^2}} \frac{ (\lambda+1)^6}{2 (\lambda-3)^3 (\lambda+5)} \right.\cr
+&\left.\frac{e^{-\frac{(\lambda+1)t}{\varepsilon^2}}}{2 (\lambda-3)^3 }(-2723 
+ 3544 \lambda - 1638 \lambda^2 + 
 288 \lambda^3 + \lambda^4)\right]
 \cr
+&\frac{45\varepsilon^6}{16(\lambda+1)^3}\left[\frac{-21659 - 172 \lambda + 1150 \lambda^2 + 148 \lambda^3 + 
 5 \lambda^4}{(\lambda+9)^2}\right.\cr
- &\left. 8e^{-\frac{4t}{\varepsilon^2}}\frac{(\lambda +1)^6 (\lambda  (5 \lambda +18)-3)}{(\lambda-3)^4 (\lambda+5)^2} +\frac{e^{-\frac{(\lambda+1)t}{\varepsilon^2}}}{(\lambda-3)^4 }(173271 - 268158 \lambda + 161953 \lambda^2  \right. \cr
 -& \left. 46276 \lambda^3 + 
 5737 \lambda^4 + 34 \lambda^5 - \lambda^6) +e^{-\frac{(\lambda+9) t}{\varepsilon ^2}} \frac{(\lambda+1)^6 }{(\lambda+5)^2 (\lambda+9)^2}
\right].\label{drabp:m6t}
\end{align}

Note that, by taking $\lambda\to 0$ we recover \eref{abp:m4}-\eref{abp:m6} obtained for ABP.

We will use the positions moments $M(2k,0,t)$ with $k=1,2,\dots$ obtained here to determine the position distribution $\rho(y,t)$ perturbatively in $\varepsilon$.

%%%%%%%%%%%%%%%%%%%%%%
\subsection{Position distribution}
Now, we look to obtain the position distribution perturbatively. For this purpose, it is important to first note that, the distribution of $\theta$, evolving by the Fokker-Planck operator $\mathcal{L}_\theta$, reaches a steady state. Thus, we can always express the solution of \eref{fp:drabp_delta} in the eigenbasis of $\mathcal{L}_\theta$ as,
\begin{subequations}
\begin{align}
P(y,\theta,t)&=\sum_{n=0}^{\infty}p_n(\theta) F_n(y,t),\\
Q(y,\theta,t)&=\sum_{n=0}^{\infty}p_n(\theta) G_n(y,t),
\end{align}
\label{pqdrabp}
\end{subequations}
where,
\bea
p_0(\theta)=\frac{1}{2\pi}\quad \text{and}\quad p_n(\theta)=\frac{1}{\pi}\cos(n\theta) \text{ for }n\geq 1
\label{drabp:pntheta}
\eea
are the eigenvalues of $\mathcal{L}_\theta$ with eigenvalue $-n^2$,
\bea
\mathcal{L}_\theta p_n(\theta)=-n^2p_n(\theta).
\eea
They obey the following orthonormality relations.
\begin{align}
\int_{-\pi}^{\pi}d\theta\,\cos(m\theta)\,p_n(\theta)=&\delta_{m,n},\\
\int_{-\pi}^{\pi}d\theta\cos\theta\,\cos(m\theta)\,p_n(\theta)=&\frac{1}{2}(\delta_{n-1,m}+\delta_{n+1,m}).
\label{drabp:pntheta_ortho}
\end{align}
Note that, integrating \eref{pqdrabp} with respect to $y$, yields,
\begin{subequations}
\bea
\int_{-\infty}^{\infty} P(y,\theta,t)\,dy=p_0(\theta),\\
\int_{-\infty}^{\infty} Q(y,\theta,t)\,dy=0,
%\frac{1}{2\pi}\sum_{n=0}^{\infty}p_n(\theta)\, e^{-(n^2+\lambda) t/\varepsilon^2}
\eea
\label{drabp:pthetat}
\end{subequations}
due to the initial condition on $\theta$, which is chosen from the stationary distribution $p_0(\theta).$
%In general, however, for specific initial condition on $\theta(0)$,
%\bea
%\int_{-\infty}^{\infty} P(y,\theta,t)\,dy=\sum_{n=0}^{\infty}p_n(\theta)\, e^{-n^2 t/\varepsilon^2},\nonumber\\
%\int_{-\infty}^{\infty} Q(y,\theta,t)\,dy=\sum_{n=0}^{\infty}p_n(\theta)\, e^{-(n^2+\lambda) t/\varepsilon^2}
%\eea
 Our aim is to obtain the marginal position distribution,
\begin{align}
\rho(y,t)\equiv\int_{-\pi}^{\pi}d\theta P(y,\theta,t)=F_0(y,t).
\end{align} 
 In general, 
 \begin{subequations} 
 \begin{align}
 F_n(y,t)&=\int_{\pi}^{\pi}d\theta\, P(y,\theta,t)\cos(n\theta),\\
 G_n(y,t)&=\int_{\pi}^{\pi}d\theta\, Q(y,\theta,t)\cos(n\theta).
 \end{align}   
 \end{subequations}
  However, the above relations are not much of a use as the functions $P(y,\theta,t)$ and $Q(y,\theta,t)$ are unknown.
Now, since \eref{fp:drabp_delta} is invariant under the transformation $(y,\varepsilon)\to(-y,-\varepsilon)$, the functions $F_n(y,t)$ and $G_n(y,t)$ also follow the same symmetry. For our initial conditions, the marginal position distribution is always symmetric about $y=0$, i.e., $F_0(y,t)=F_0(-y,t)$. Therefore $F_0(y,t)$ is an even function of $\varepsilon$.

 Putting Eqs.~\eqref{pqdrabp} in \eref{fp:drabp_delta} and thereafter using the orthonormality relations for $p_n(\theta)$, we obtain PDEs for $F_m$ and $G_m$. For $m=0$,
\begin{subequations}
\label{drabp:pdeA0B0}
\begin{align}
\varepsilon^2 \frac{\partial F_0}{\partial t}&=-\varepsilon \sqrt{2(\lambda+1)}\,  \frac{\partial G_1}{\partial y},\\
\left[\varepsilon^2 \frac{\partial }{\partial t}+\lambda\right]G_0&=-\varepsilon \sqrt{2(\lambda+1)} \, \frac{\partial F_1}{\partial y},
\end{align}
\end{subequations}
and for all other $m>0$,
\begin{subequations}
\label{drabp:pdeAmBm}
\begin{align}
\left[\varepsilon^2 \frac{\partial }{\partial t}+m^2\right]F_m&=-\varepsilon\sqrt{\frac{\lambda+1}{2}}\left( \frac{\partial G_{m+1}}{\partial y}+\frac{\partial G_{m-1}}{\partial y} \right),\\
\left[\varepsilon^2 \frac{\partial }{\partial t}+m^2+\lambda\right]G_m&=-\varepsilon\sqrt{\frac{\lambda+1}{2}}\left( \frac{\partial F_{m+1}}{\partial y}+\frac{\partial F_{m-1}}{\partial y} \right).
\end{align}
\end{subequations}

From \eref{drabp:pthetat}, it is clear that we do not have a series in $\nu$, so we straightaway look for series solution of $F_m$ and $G_m$ in the form,
\begin{subequations}
\begin{align}
F_m(y,t)&=\sum_{k=0}^{\infty}\varepsilon^k A_m^k(y,t),\label{drabp:am}\\
G_m(y,t)&=\sum_{k=0}^{\infty}\varepsilon^k B_m^k(y,t).
\label{drabp:bm}
\end{align}
\end{subequations}
By definition, $A_m^k=B_m^k=0$ for $k<0$. Moreover, since $A_0(y,t)$ is an even function of $\varepsilon,$ we must have $A_0^k=0$ for odd integers $k$, i.e., $A_0^1=A_0^3=\dotsb =0$. Therefore,
\begin{align}
A_0(y,t)=\sum_{k=0}^{\infty}\varepsilon^{2k} A_0^{2k}(y,t).
\label{drabp:marginal}
\end{align}
The symmetry, $\int_{-\pi}^{\pi}d\theta\,P_\sigma(y,\theta,t)=\int_{-\pi}^{\pi}d\theta\,P_{-\sigma}(-y,\theta,t)$, implies that $\int_{-\pi}^{\pi}d\theta\,Q(y,\theta,t)=-\int_{-\pi}^{\pi}Q(-y,\theta,t)$. Therefore $B_0(y,t)$ should also be an odd function of $\varepsilon$,
\begin{align}
B_0(y,t)=\sum_{k=0}^{\infty}\varepsilon^{2k+1} B_0^{2k+1}(y,t).
\end{align}

Now, putting Eqs.~\eqref{drabp:am} and \eqref{drabp:bm} in \eref{drabp:pdeA0B0} and comparing powers of $\varepsilon$,
\begin{subequations}
\begin{align}
\frac{\partial A_0^{k-2}}{\partial t}&=-\sqrt{2(\lambda+1)}\frac{\partial B_1^{k-1}}{\partial y},\label{drabp:a0k}\\
\frac{\partial B_0^{k-2}}{\partial t}+\lambda B_0^k&=-\sqrt{2(\lambda+1)}\frac{\partial A_1^{k-1}}{\partial y}.\label{drabp:b0k}
\end{align}\label{drabpaokbok}
\end{subequations}
On the other hand, for $m\geq 1$, we get from \eref{drabp:pdeAmBm},
\begin{subequations}
\begin{align}
m^2A_m^k+\frac{\partial A_m^{k-2}}{\partial t}&=-\sqrt{\frac{\lambda+1}{2}}\,\frac{\partial}{\partial y}\Big[B_{m-1}^{k-1}+ B_{m+1}^{k-1}\Big],\label{drabp:amk}\\
(m^2+\lambda)B_m^k+\frac{\partial B_m^{k-2}}{\partial t}&=-\sqrt{\frac{\lambda+1}{2}}\,\frac{\partial}{\partial y}\Big[A_{m-1}^{k-1}+ A_{m+1}^{k-1}\Big].\label{drabp:bmk}
\end{align}
\label{drabpamkbmk}
\end{subequations}

Before finding solutions for $A_m^k$ and $B_m^k$, let us first simplify the series in Eqs.~\eqref{drabp:am} and \eqref{drabp:bm}. Putting $k=0$ in \eref{drabp:amk},
we have $m^2A_m^0=0 $. Thus,
\begin{align}
A_m^0=\delta_{m,0}A^0_0,
\label{drabp:am0}
\end{align} 
i.e., $k=0$ term in the series \eref{drabp:am} exists for $m=0$ only. 

Putting $k=0$ in \eref{drabp:b0k} and \eqref{drabp:bmk}, we have $B_m^0=0$ for all $m\geq 0.$ 
Using this fact and putting $k=1$ in  \eref{drabp:amk}, we have, $A^1_m=0$ for all $m\geq 1$. Note that, we already have $A_0^1=0$. Again, putting $k=1$ in \eref{drabp:bmk}, we get 
\begin{align}
(m^2+\lambda)B^1_m=-\sqrt{\frac{\lambda+1}{2}}\left(\frac{\partial A_{m-1}^0}{\partial y}+\frac{\partial A_{m+1}^0}{\partial y}\right).
\end{align}
Further, using \eref{drabp:am0}, and combining with the earlier result for $m=0$, we get,
\begin{align}
B^1_m=-\delta_{m,1}\frac{1}{\sqrt{2(\lambda+1)}}\,\frac{\partial A_0^0}{\partial y} + \delta_{m,0} B_{0}^1.
\label{b1m}
\end{align}

\begin{figure}
\centering\includegraphics[width=0.7\hsize]{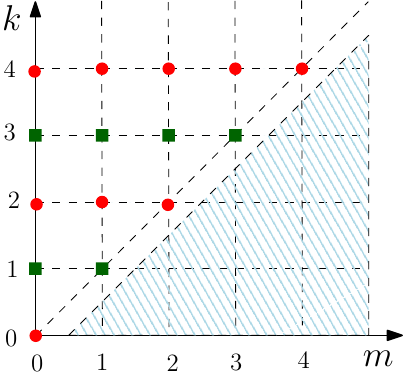}
\caption{Graphical representation of recursive determination of non-vanishing  $A_m^k$ (red dots) and $B_m^k$ (blue crosses) following the Eqs.~\eqref{drabp:a0k}-\eqref{drabp:bmk}.}\label{fig:AmBm}
\end{figure}

One can systematically proceed by putting $k=2,3, 4,\dotsc$ and obtain 
non-vanishing coefficients $A_m^k$ and $B_m^k.$ This process is best illustrated graphically  on the $m$-$k$ plane (see Fig.~\ref{fig:AmBm}). Since, $A_m^k$ only depends on $A_m^{k-2}$ and $B_{m\pm 1}^{k-1}$ and $B_m^k$ also follow a similar pattern, it is clear that $A_m^k$ are non-zero on even $k$ lines while $B_m^k$ are non-zero on odd $k$ lines only. Moreover, both $A_m^k$ and $B_m^k$ vanish on the lower triangle $m >k$. Therefore,  Eqs.~\eqref{drabp:am}-\eqref{drabp:bm} can be refined to,
\begin{subequations}
 \begin{align}
A_m&=\sum_{k=m}^{\infty}\varepsilon^{2k}A^{2k}_m,\\
B_m&=\sum_{k=m}^{\infty}\varepsilon^{2k+1}B^{2k+1}_m.
\end{align}
\end{subequations}

%%%%%%%%%%%%%%%%%%%%%
%%%%%%%%%%%%%%%%%%%%%

Using the above series expansion, we now proceed to compute the marginal distribution $A_0(y,t)$ perturbatively.
The leading order term $A_0^0$ satisfies
\begin{align}
\frac{\partial A_0^{0}}{\partial t}&=-\sqrt{2(\lambda+1)}\frac{\partial B_1^{1}}{\partial y}.
\label{a00pde:drabp}
\end{align}
which is obtained by putting $k=2$ in \eref{drabp:a0k}. Now, $B^1_1$ can be obtained by putting $k=1$, $m=1$ in \eref{drabp:bmk},
\begin{align}
B^1_1=-\frac{1}{\sqrt{2(\lambda+1)}}\frac{\partial A_0^0}{\partial y}.
\end{align}
Inserting this back in \eref{a00pde:drabp} we get the diffusion equation,
\begin{align}
\frac{\partial A_{0}^{0}}{\partial t}=\frac{\partial^2 A_{0}^{0}}{\partial y^2}.
\end{align}
Let us remark that, $A_0^0(y,t)$ should have the same properties as a normalized probability density function since $[A_0(y,t)/A_0^0(y,t)] \to  1$ as $t \to \infty$. This also demands that, $\int dy\,A_0^k(y,t)=0$ for $k>0$. Hence, the above equation  can immediately be solved to obtain a Gaussian distribution,
\begin{align}
A_0^0(y,t)=\frac{e^{-y^2/(4t)}}{\sqrt{4\pi t}}.
\label{drabp:a0_0}
\end{align}

The subsequent coefficients $A_0^k(y,t)$ provide systematic correction to this Gaussian form. Noting that $\varepsilon^2$ has the dimension of time $t$, we expect the following diffusive scaling form,
\begin{align}
A_0^{2k}(y,t)=\frac{e^{-\frac{y^2}{4t}}}{\sqrt{4\pi t}} \frac 1{t^k}\, q_{2k}\Bigl(\frac{y}{\sqrt{4t}}\Bigr).
\label{drabp:q2kgen}
\end{align}

For the next order correction, putting $k=4$ in \eref{drabp:a0k}, we get,
\begin{align}
\frac{\partial A_0^{2}}{\partial t}&=-\sqrt{2(\lambda+1)}\frac{\partial B_1^{3}}{\partial y}.
\label{a02pde:drabp}
\end{align}
Putting $k=3$, $m=1$ in \eref{drabp:bmk} to get $B_1^3$,
\begin{align}
(\lambda+1)B_1^3=-\frac{\partial B_1^1}{\partial t}-\sqrt{\frac{\lambda+1}{2}}\left( \frac{\partial A_0^2}{\partial y}+\frac{\partial A_2^2}{\partial y}\right). 
\end{align}

\begin{figure}
\fl
\centering\includegraphics[width=1.\hsize]{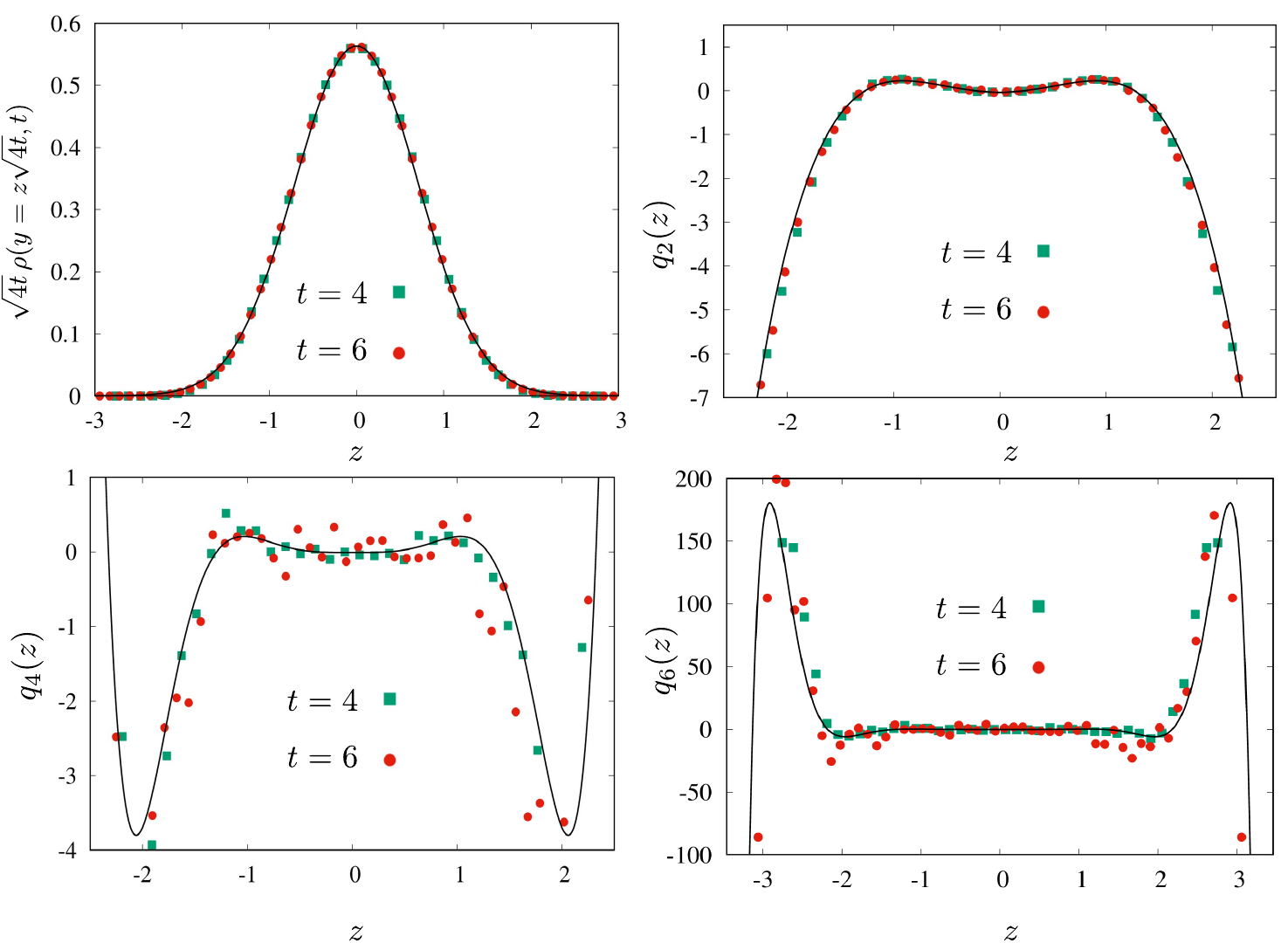}
\caption{Plots for the long-time distribution of a DRABP with $D_R=2.0$, $\gamma=1$, at two different times---(a) compares the leading order contribution of the scaled Gaussian distribution \eref{drabp:a0_0} with that obtained from numerical simulation; (b), (c) and (d) compare $q_2(z)$, $q_4(z)$ and $q_6(z)$ obtained in \eref{drabp:q2}, \eref{app:drabpq4_1} and \eref{app:drabpq6_1} with the same obtained from numerical simulations. In all the plots the black solid lines denote the theoretical predictions, while the numerical simulations are indicated by colored symbols.}
\label{F:drabp}
\end{figure}

Thereafter, using $k=2,m=2$ in \eref{drabp:amk}, to obtain $A_2^2$, we get an inhomogeneous diffusion equation for $A_0^2$,
\begin{align}
\Bigg[\frac{\partial }{\partial t}-\frac{\partial^2 }{\partial y^2}\Bigg]A_0^{2}=-\frac{\partial^2}{\partial y^2}\left(\frac{1}{\lambda+1}\frac{\partial }{\partial t}-\frac{1}{8}\frac{\partial^2 }{\partial y^2}\right)A_0^0,
\end{align}
as mentioned in \eref{strategy:inhomo_diff}. To solve the above equation, we anticipate the following scaling form for $A_0^2(y,t)$,
\begin{align}
A_0^2(y,t)=\frac{e^{-y^2/(4t)}}{\sqrt{4\pi t}}\frac{1}{t}q_2\left(\frac{y}{\sqrt{4t}}\right).
\end{align}
Infact, in general, for higher orders,
\begin{align}
A_0^{2k}(y,t)=\frac{e^{-y^2/(4t)}}{\sqrt{4\pi t}}\frac{1}{t^k}q_{2k}\left(\frac{y}{\sqrt{4t}}\right).
\end{align}
Using the above mentioned scaling form and the expression for $A_0^0(y,t)$, we get for $q_2(z)$,
\begin{align}
q_2''(z)-2zq_2'(z)+4q_2(z)=-\frac{(\lambda -7)}{8 (\lambda +1)}(3-12z^2+4z^4).
\end{align}
This leads to the general solution for $q_2(z)$ using \eref{eq:qn_gen},
\begin{align}
q_2(z)=c_2 H_2(z)+\frac{(\lambda -7)}{16 (\lambda +1)} \left(-3 z^2+2 z^4\right).
\end{align}

The normalization condition $\int_{-\infty}^{\infty}dz A_0(y,t)$ is satisfied trivially for arbitrary values of $C_2$ and thus as mentioned before we take recourse to the moments to evaluate $C_2$.  In fact, at each order the constant $C_{2k}$ is determined by comparing the coefficient of $(\tau/t)^k$ of $M(2k,0,t)/(4t)^k$ obtained from the two methods: the exact computation in Sec.~\ref{sec:drabpmoments} and using the series \eref{drabp:marginal}; where the latter is simply given by,
\begin{align}
\int_{-\infty}^{\infty}dz\, z^{2k} q_{2k}(z)e^{-z^2}/\sqrt{\pi}.
\label{drabp:m2k_series}
\end{align}
Following this procedure for $k=1$, we get $C_2=\frac{5-3 \lambda }{64 (\lambda +1)}$, which leads to,
\begin{align}
q_2(z)=\frac{3 \lambda -5}{32 (\lambda +1)}+\frac{(13-3 \lambda ) }{8 (\lambda +1)}z^2+\frac{(\lambda -7) }{8 (\lambda +1)}z^4.
\end{align}

Similarly we can find the subleading contributions $A_0^4(y,t)$ and $A_0^6(y,t)$ (see \ref{app:drabpdetails}). They satisfy the inhomogeneous differential equation, 
\begin{align}
\left[\frac{\partial}{\partial t}-\frac{\partial^2 }{\partial y^2}\right] A_0^{2k}(y,t)=S_{2k}(y,t),
\label{drabp:q2}
\end{align}
as announced in \eqref{strategy:inhomo_diff}. The explicit forms of $S_4$ and $S_6$ are given in the appendix. Substituting the ansatz,
\begin{align}
A_0^{2k}(y,t)=\frac{e^{-y^2/(4t)}}{\sqrt{4\pi t}}\frac{1}{t^k}q_{2k}\left(\frac{y}{\sqrt{4t}}\right),
\end{align}
leads to inhomogeneous Hermite equation \eref{gen:hermite_inh} for $q_{4}(z)$ and $q_{6}(z)$, whose general solution is given by \eqref{eq:qn_gen}. We find the explicit solutions (see Appendix) as,
\begin{align}
q_{2k}(z)=\sum_{n=0}^{2k}\alpha_{2k,n}(\lambda)\,z^{2n},
\end{align}  
where $\alpha_{2k,n}(\lambda)$ is a polynomial in $\lambda$. The expressions of $\{\alpha_{4,n};0\leq n\leq 4\}$ and $\{\alpha_{6,n};0\leq n\leq 6\}$ are rather long and are given in \ref{app:drabpdetails}. Figure \ref{F:drabp} compares the leading order Gaussian
along with the subleading corrections $q_2(z)$ , $q_4(z)$ and $q_6(z)$ with the same extracted from the numerical simulations of DRABP and shows reasonably good agreement.

\section{Conclusion}\label{conclusion}

In summary, we develop a unifying framework to systematically study the position distribution $\rho(x,t)$ of active particles at late times $t$ much larger than the persistence time $\tau_0$. In this regime, the position distribution admits a perturbative series in powers of $\tau_0/t$. Using the examples of four well-known active particle models, namely, the run-and-tumble particle, the active Ornstein-Uhlenbeck particle, the active Brownian particle and the direction-reversing active Brownian particle, we show that the leading term generically satisfies the diffusion equation with an effective diffusion coefficient $D_{\text{eff}}$ that depends on the specific model. We further find that the higher-order subleading corrections, again, generically satisfy an inhomogeneous diffusion equation, where the source term is model dependent and involves the previous order solutions. Consequently, the higher-order corrections also admit diffusive scaling. The distribution of the scaled position $z=x/\sqrt{4D_{\text{eff}}t}$ can be generically written as $q(z,t)=(e^{-z^2}/\sqrt{\pi})\sum_{k=0}^{\infty}(\tau_0/t)^k\, q_{2k}(z)$, where $q_{2k}(z)$ is a $4k$-th order polynomial in $z$ that satisfies an inhomogeneous Hermite differential equation.

The most prominent signatures of activity are encoded in the large deviation function associated with the $x\sim O(t)$ fluctuations. However, these rare events are hard to access experimentally, and one is often limited to the events of  $O(\sqrt{t})$ fluctuations.  
 The higher-order non-Gaussian corrections, obtained here, highlight the signature of activity at the scale $x\sim O(\sqrt{t})$. These corrections cannot be obtained trivially by expanding the large deviation function beyond quadratic order, as the subleading corrections to the large deviation form of the probability distribution are also important---as shown explicitly, here, for the ABP case.

The quantitative predictions for the deviation of the position distribution from Gaussian should be verifiable in experiments, as the $O(\sqrt{t})$ fluctuations are easily accessible. In fact, our framework provides a quantitative test for the suitability of a particular model for describing a given active system.

For simplicity, here we have restricted ourselves to one-dimension. It is, however, straightforward to generalize the procedure to higher dimensions. In fact, our framework is quite generic and expected to be applicable to any Fokker-Planck or master equation involving a small perturbative parameter. For example it would be interesting to study the signatures of activity in other variants of active particle models~\cite{Santra2020,Shee2021_speedfluctuation,Grossmann2016,Basu2020,Chakrabarti2021}
It would be also interesting to ask similar questions for the generalized run-and-tumble process, where the large time scaling of the position fluctuations is anomalous~\cite{Dean2021}. Another future direction is to extend the framework to systematically study the subleading corrections to the long-time $t^{-3/2}$ behavior of the first-passage time probability distributions of active motions. 
\section{Acknowledgments}
U.B. acknowledges support from the
Science and Engineering Research Board (SERB), India,
under a Ramanujan Fellowship (Grant No. SB/S2/RJN-077/2018).
\appendix
\addtocontents{toc}{\fixappendix}
\section{Wronskian for a general  $n$ of the Hermite differential equation}\label{App:wronskian}
The Hermite differential equation,
\begin{align}
q_n''(z)-2zq_n'(z)+4nq_n(z)=0,
\end{align}
has two linearly independent roots given by,
\begin{align}
u_{1,n}(z)=H_{2n}(z)\qquad\text{and}\qquad u_{2,n}(z)=z\, {}_1F_1\left(\frac 1 2-n,\frac 3 2,z^2\right).
\end{align}
The corresponding Wronskian, given by,
\begin{align}
W_n(z)=u_{1,n}(z)u_{2,n}'(z)-u_{1,n}'(z)u_{2,n},
\end{align}
satisfies the differential equation,
\begin{align}
W_n'(z)-2zW_n(z)=0.
\end{align}
Thus, we have $W_n(z)=\omega_0~e^{z^2}$, where the constant $\omega_0$ can be evaluated from the initial condition $W_n(0)=(-1)^n(2n)!/n!$. Finally, we get,
\begin{align}
W_n(z)=(-1)^n\frac{(2n)!}{n!}e^{z^2}.
\end{align}
This result has been quoted in the main text.
%%%%%%%%%%%%%%%%%%%%%%%%%%%%%%%%%%%%%%%%%%
\section{Derivation of the relation \eref{rtpmomentcompare}}
We prove \eref{rtpmomentcompare} for RTP in this appendix. Multiplying both sides of \eref{rtp:inhomo} by $z^{2l}\,e^{-z^2}$, and integrating with respect to $z$, we get,
\bea
4(n-l)\int_{-\infty}^{\infty}dz\,e^{-z^2} z^{2l}q_{2n}(z)+2l(2l-1)\int_{-\infty}^{\infty}dz\,e^{-z^2}z^{2l-2}q_{2n}(z)\cr
=4(n-l)(n-l-1)\int_{-\infty}^{\infty}dz\,e^{-z^2}z^{2l}q_{2n-2}(z),
\eea
for arbitrary $(n,l)$. Setting $l=k+1=n$ yields,
\bea
\int_{-\infty}^{\infty} \frac{dz}{\sqrt{\pi}}~ e^{-z^2}  z^{2k}\, q_{2(k+1)}(z)=0.
\label{app:rtp1}
\eea
Setting $l=k+1=n-1$ and using \eref{app:rtp1} we have,
\bea
\int_{-\infty}^{\infty} \frac{dz}{\sqrt{\pi}}~ e^{-z^2}  z^{2k}\, q_{2(k+2)}(z)=0.
\label{app:rtp2}
\eea
Next, setting $l=k+1=n-2$ and using the above two relations, we get,
\bea
\int_{-\infty}^{\infty} \frac{dz}{\sqrt{\pi}}~ e^{-z^2}  z^{2k}\, q_{2(k+3)}(z)=0.
\label{app:rtp3}
\eea
We can proceed similarly, by setting $l=k+1=n-3,\,n-4,\dotsc$ and show that,
\bea
 \int_{-\infty}^{\infty} \frac{dz}{\sqrt{\pi}}~ e^{-z^2}  z^{2k}\, q_{2n}(z)=0, \qquad\text{for }n>k
 \label{app_b:rtp}
\eea
as announced in \eref{rtpmomentcompare}.

\section{Extracting the higher order corrections to Gaussian from the exact solution of RTP}\label{app:rtp_exact}

 In this section, we verify the results obtained using our perturbative framework with the exact distribution known from earlier studies. The exact solution of the Telegraphers equation \eref{eq:rtp_FPx} for the initial condition $\sigma(0)=\pm 1$ with equal probability $1/2$, is given by~\cite{Malakar2017},
\begin{align}
 P(x,t)&=\frac{e^{-\gamma t}}{2}\left[\delta(x-v_0t)+\delta(x+v_0t)\right]\cr 
 &+\frac{e^{-\gamma t}}{2}\left[I_0(\gamma t\sqrt{1-(x/v_0 t)^2})+\frac{I_1(\gamma t\sqrt{1-(x/v_0 t)^2})}{\sqrt{1-(x/v_0 t)^2}}\right]\Theta(v_0t-|x|),
 \label{rtp-exact}
\end{align}
where $I_\nu(w)$ is the modified Bessel function of order $\nu$. For $t\gg\tau$, the weight of the boundary $\delta$-functions, characterizing the ballistic spread, vanishes and the Heaviside-$\Theta$ function becomes unity for $x\sim O(\sqrt{t})$. In this regime, substituting $v_0=\sqrt{D_\rt/\tau}$, we can expand $P(x,t)$ as a power series in $\tau/t$ as,
\begin{equation}
P(x,t)=\frac{1}{\sqrt{4\pi  D_\rt t}}\exp\left(-\frac{x^2}{4 D_\rt t}\right) \sum_{n=0}^{\infty}\left(\frac{\tau_0}{t}\right)^n q_n\left(\frac{x}{\sqrt{4D_\rt t}}\right) +O(e^{-t/\tau}).
\label{eq:rtpqn}
\end{equation}
Using the series expansion of $I_0(w)$ and $I_1(w)$ for large $w$, the first few terms in above expansion can be obtained as,
\begin{subequations} 
\begin{align}
q_0(z)&=1,\label{rtpq0}\\
q_2(z)&=-\frac{1}{4}\left(1-8z^2+4z^4\right),\label{eq:q1exact}\\
q_4(z)&=-\frac{1}{32}\left(3 + 48 z^2 - 216 z^4 + 128 z^6 - 16 z^8\right),\label{eq:q2exact}\\
q_6(z)&=-\frac{1}{384}\left(45 + 360 z^2 + 2700 z^4 - 9600 z^6 + 6000 z^8 - 1152 z^{10} + 64 z^{12}\right)\label{eq:q3exact}.
\end{align}
\label{rtp_qs}
\end{subequations}
These match exactly with the ones obtained using our perturbative framework, namely, \eref{rtp:rho0}, \eref{rtp:q2}, \eref{rtp:q4} and \eref{rtp:q6} in the main text, thus validating our procedure.

%%%%%%%%%%%%%%%%%%%%%%%%%%%%%%%%%%%%%%%%%%
\section{Extracting the higher order corrections to Gaussian from the exact solution of AOUP}\label{App:aoup}
 In this appendix we illustrate the validity of the perturbative procedure for the AOUP discussed in Sec.~\ref{Sec:AOUP}. In particular, using the exact expression of the joint probability distribution $P(y,u,t)$ in \eref{hnbasis}, we explicitly calculate $F_n(y,t)$ for a few $n$ using \eref{under:F_n} and demonstrate that they satisfy \eref{under:am}. Subsequently, we also calculate  $F_{n,l}(y,t)$, defined by \eqref{u:aml}, for a few $(n,l)$ and demonstrate that they satisfy \eref{under:aml}. Finally, we show that the corrections to the Gaussian distribution obtained from out perturbative technique is consistent with those extracted from the exact solution. 

We begin by rewriting the Langevin equations \eref{under:langevin1} and \eref{under:langevin2} in terms of the scaled variables $u=v\sqrt{\tau/D}$ and $y= x/\sqrt{D}$,
\bea
 \dot{y}(t)=\frac{1}{\varepsilon}\,u(t),\qquad\qquad \dot{u}(t)=-\frac{1}{\varepsilon^2} u(t)+\frac{1}{\varepsilon}\,\eta(t),
 \label{langevin:aoup}
 \eea
where $\varepsilon=\sqrt{\tau}$. We consider the initial condition $y(0)=0$ and $u(0)=0$, for which the mean $\la y(t)\ra=0=\la u(t)\ra$ for all $t$. Note that, the active Ornstein-Uhlenbeck process \eref{langevin:aoup} is linear in the Gaussian white noise $\eta(t)$, and thus, the joint distribution $P(y,u,t)$ is given by the bi-variate Gaussian, 
 \bea
 P(y,u,t)=\frac{1}{2\pi\sqrt{\det\Sigma}}\exp\left[-\frac{1}{2}X^T \Sigma^{-1}X\right].
 \label{under:joint}
 \eea
 
Here, $X^T=\begin{bmatrix}
y & u
\end{bmatrix}$ and $\Sigma$ is the covariance matrix, given by
\bea
\Sigma=\begin{bmatrix}
\la y^2(t)\ra & \la y(t)u(t)\ra\\
\la y(t)u(t)\ra & \la u^2(t)\ra
\end{bmatrix},
\eea 
with 
\begin{subequations}
\bea
\la y^2(t)\ra=2t-\varepsilon^2(3-\nu)(1-\nu),\label{u:y2t}\\
 \la u^2(t)\ra=(1-\nu^2),\\
 \la y(t)u(t)\ra=\varepsilon(1-\nu)^2,
\eea
\end{subequations}
where $\nu=e^{-t/\varepsilon^2}$. 
The marginal position distribution is clearly a Gaussian,
\bea
F_0(y,t)=\int_{-\infty}^{\infty}P(y,u,t)\,du=\frac{e^{-\frac{y^2}{2\la y^2(t)\ra}}}{ \sqrt{2\pi\la y^2(t)\ra}},
\eea
where $\la y^2(t)\ra$ is given by \eref{u:y2t}. 
We can readily see that $F_0(y,t)$ has a power series in $\nu$,
\bea
F_0(y,t)=\frac{e^{-\frac{y^2}{2t(2-3 \varepsilon ^2/t)}}}{\sqrt{2\pi t(2 -3  \varepsilon ^2/t)}}\left[1-\frac{(\varepsilon ^2/t) \left(2 -y^2/t-3 \varepsilon ^2/t\right)}{(2  -3  \varepsilon ^2/t)^2}\nu+O(\nu^2)\right].
\eea
Thus, according to \eref{u:aml} we have,
\bea
F_{0,0}(y,t)=\frac{e^{-\frac{y^2}{2t(2-3 \varepsilon ^2/t)}}}{\sqrt{2\pi t(2 -3  \varepsilon ^2/t)}},\label{under:f00}\\
F_{0,1}(y,t)=-\frac{e^{-\frac{y^2}{2t(2-3 \varepsilon ^2/t)}}}{\sqrt{2\pi t(2 -3  \varepsilon ^2/t)}}\frac{(\varepsilon ^2/t) \left(2 -y^2/t-3 \varepsilon ^2/t\right)}{(2  -3  \varepsilon ^2/t)^2},
\eea
and so on.  

Next, we calculate $F_1(y,t)$ using \eref{under:F_n} as,
\bea
F_1(y,t)=-\frac{ \exp(-\frac{y^2}{4 t-2 (3-\nu) (1-\nu) \varepsilon ^2})  y \varepsilon(1-\nu)^2}{\sqrt{\pi } \Big[2 t-(3-\nu) (1-\nu) \varepsilon ^2\Big]^{3/2}}.
\eea
Now expanding the above expression as a power series in $\nu$ and using \eref{u:aml} we get,
\bea
F_{1,0}(y,t)=\frac{\,e^{-\frac{y^2}{2t(2-3 \varepsilon ^2/t)}}}{\sqrt{\pi t(2 -3  \varepsilon ^2/t)^3}}\frac{y\varepsilon}{t},
\\
F_{1,1}(y,t)=-\frac{ e^{-\frac{y^2}{2t(2-3 \varepsilon ^2/t)}}}{\sqrt{\pi t(2 -3  \varepsilon ^2/t)^7}}\frac{y\varepsilon}{t}\left(4-6\varepsilon ^2/t -y^2\varepsilon^2/t^2\right).
\eea

Similarly, we calculate $F_2(y,t)$ using \eref{under:F_n} and expanding it as a power series in $\nu$ and using \eref{u:aml}, we obtain,
\bea
F_{2,0}(y,t)&=&-\frac{2e^{-\frac{y^2}{2t(2-3 \varepsilon ^2/t)}}}{\sqrt{2\pi t(2 -3  \varepsilon ^2/t)^5}}\frac{\varepsilon^2}{t}\left( 2-3\varepsilon^2/t-y^2/t \right),\\
F_{2,1}(y,t)&=&\frac{4e^{-\frac{y^2}{2t(2-3 \varepsilon ^2/t)}}}{\sqrt{2\pi t(2 -3  \varepsilon ^2/t)^9}}\frac{\varepsilon^2}{t}\left( 16-27(\varepsilon^2/t)^3+72(\varepsilon^2/t)^2  \right.\cr
&& \left. -60(\varepsilon^2/t) +12(y^2\varepsilon^2/t^2)+(y^4\varepsilon^2/t^3)\right).
\eea

At this point, we can readily verify \eref{under:aml} for $(n,l)=(0,0)$, $(0,1)$, $(1,0)$, and $(1,1)$.

Now, the perturbative corrections to the Gaussian in the long time limit can be obtained by expanding \eref{under:f00} as a power series in $\varepsilon$, 
\begin{align}
F_{0,0}(y,t)&=\frac{e^{-\frac{y^2}{4 t}}}{\sqrt{4\pi t }}\left(1-\frac{ \left(3 y^2-6 t\right)}{8 }\left(\frac{\varepsilon}{t}\right)^2+\frac{9 \left(12 t^2-12 t y^2+y^4\right)}{128 } \left(\frac{\varepsilon}{t}\right)^4\right.
\cr
&\left.+\frac{9 (120 t^3 - 180 t^2 y^2 + 30 t y^4 - y^6)}{1024}\left(\frac{\varepsilon}{t}\right)^6+O\left[ \left(\frac{\varepsilon}{t}\right)^8 \right]  \right).
\end{align}
Taking the scaling $y=z\sqrt{4t}$, we get $F_{0,0}(y,t)=f_{0,0}(z,t)/\sqrt{4t}$,
\begin{align}
f_{0,0}(z,t)=&\frac{e^{-z^2}}{\sqrt{\pi}}\left( 1+\frac{3\left(1-2 z^2\right)}{4}\left(\frac{\varepsilon^2}{t}\right)^2+\frac{9\left(3-12 z^2+4z^4\right)}{32} \left(\frac{\varepsilon^2}{t}\right)^2\right.\cr
&\left.+\frac{9(15-90z^2+60z^4-8z^6)}{128}\left(\frac{\varepsilon^2}{t}\right)^3
+O\left[\left(\frac{\varepsilon^2}{t}\right)^3\right]\right)\cr
=&\frac{e^{-z^2}}{\sqrt{\pi}}\Bigg( 1-\frac{3}{8}H_{2}(z)+\frac{9}{128}H_4(z)-\frac{9}{1024}H_6(z)+O\left[\left(\frac{\varepsilon^2}{t}\right)^3\right]\Bigg)
\end{align}
This is consistent with the corrections $q_2(z),$ $q_4(z)$ obtained using our perturbative strategy in the main text.
%%%%%%%%%%%%%%%%%%%%%%%%%%%%%%%%%%%%%%%%%%%%%%%%%%
\section{Intermediate steps in the calculation of $A_0^6(y,t)$ for ABP}\label{Sec:ABPa06}
In this section, we provide the intermediate steps leading to \eref{abp:q6final} starting from \eref{abp:a06eq}. The subleading contribution $A_0^6(y,t)$ satisfies the inhomogeneous diffusion equation \eref{abp:a06eq}, where the inhomogeneous part $S_6(y,t)$ is given by,
%\begin{align} 
\bea
\fl S_6(y,t)=\frac{\partial^2 }{\partial y^2}\left[\Bigg(-\frac{\partial}{\partial t}+\frac 18 \frac{\partial^2}{\partial y^2}\Bigg)A_0^4(y,t)+\Bigg(\frac{\partial^2}{\partial t^2}-\frac{9}{32}\frac{\partial^3}{\partial y^2 \partial t}+\frac{5}{288}\frac{\partial^4}{\partial y^4}\Bigg)A_0^2(y,t) \right.\cr
+\left.\Bigg(
-\frac{\partial^3}{\partial t^3}+\frac{57}{128}\frac{\partial^4}{\partial y^2 t^2}-\frac{307}{5184}\frac{\partial^5}{\partial y^4 \partial t}+\frac{401}{165888}\frac{\partial^6}{\partial y^6}\Bigg)A_0^0(y,t) \right].
\eea
%\end{align}
Using the scaling ansatz for $A_0^6(y,t)$ and the explicit forms of $A_0^{0,2,4}(y,t)$, we get,
\begin{align}
s_6(z)=-&\frac{162575}{49152}+\frac{36295}{12288}z^2+\frac{4010615}{12288}z^4-\frac{2280089}{4608}z^6+\frac{1895057}{9216}z^8\cr
&-\frac{68831}{2304}z^{10}+\frac{343}{256}z^{12}.
\end{align}
The general solution for $q_6(z)$ can then be obtained in terms of an arbitrary constant $C_6$ using \eref{eq:qn_gen}
\begin{align}
q_6(z)=C_6H_6(z)&-\frac{162575}{49152}+\frac{36295}{12288} z^2+\frac{4010615}{12288} z^4-\frac{2280089}{4608} z^6+\frac{1895057 }{9216}z^8\cr
&-\frac{68831 }{2304}z^{10}+\frac{343}{256} z^{12}.
\end{align}
The constant $C_6$ can again be found by comparing the coefficient of $(\tau/t)^3$ in the expansion of $M(6,0,t)/(4t)^3$ obtained using the distribution \eref{abp:marginal} to that obtained from \eref{abp:m6}. This procedure yields $C_6=60595/42467328$, using which we finally get,
\begin{align}
q_6(z)=&-\frac{302975}{1769472}-\frac{92375 }{147456}z^2 +\frac{293635 }{442368}z^4+\frac{1790095 }{165888}z^6 -\frac{325727 }{36864}z^8\cr
&+\frac{17437 }{9216}z^{10}-\frac{343 }{3072}z^{12}.
\end{align}
This is the result quoted in \eref{abp:q6final} in the main text. Putting the above form in
\begin{align}
A_0^{6}(y,t)=\frac{1}{t^{3}}\,q_{6}\left(\frac{y}{\sqrt{4t}}\right)\frac{e^{-y^2/(4t)}}{\sqrt{4\pi t}},
\end{align}
 completely determines the $O(\varepsilon^6)$ contribution to the position distribution.

%
%The inhomogeneous term is given by,
%\begin{align}
%s_6(z)=-&\frac{162575}{49152}+\frac{36295}{12288}z^2+\frac{4010615}{12288}z^4-\frac{2280089}{4608}z^6+\frac{1895057}{9216}z^8\cr
%&-\frac{68831}{2304}z^{10}+\frac{343}{256}z^{12}.
%\end{align}
%Equation \eref{abp:q6_eq} can be solved using \eref{eq:qn_gen} to obtain a general solution in terms of an arbitrary constant $C_6$, 

%%%%%%%%%%%%%%%%%%%%%%%%%%%%%%%%%%%%%%%%%%%%%%%%%%%%%%%%
\section{Extracting the higher order corrections to Gaussian for ABP using Mathieu equations}\label{mathieu}

 The subleading contributions to position distribution of an ABP can also  be extracted from the exact solution of the corresponding Mathieu equation.
In this appendix we extract these contributions explicitly and show that they agree with the same obtained using the perturbative procedure in Sec.~\ref{Sec:ABP}.

Basu et. al. in their work \cite{Basu2019} studied the generating function of an ABP,
\begin{equation}
Q_p(u,t)=\la e^{-pv_0\int_0^t d\tau \cos\theta(\tau)} \ra
\end{equation}
with $\theta(0)=u$.
 Using a backward Feynman-Kac equation for $Q_p(u,t)$, they computed the exact generating function as,
\begin{align}
Q_p(u,t)=\sum_{n=0}^{\infty}{\cal A}_{2n} \ce_{2n}\left( \frac{u}{2},\frac{2p}{D_R} \right) \exp\left[-\frac{t D_R}{4}\, a_{2n}\left(\frac{2p}{D_R}\right)\right].
\label{abp:qput}
\end{align} 
 Here $\ce_{2n}(\nu,q)$ are solutions of the Mathieu equation,
 \begin{equation}
 \psi''(v)+\left(a-2q\cos(2v)\right)\psi(v)=0,
 \end{equation}
 which are $\pi$-periodic and even in $\nu$, with eigenvalues $a=a_{2n}(q)$. The coefficient ${\cal A}_{2n}$ can be determined from the initial condition,
 \begin{equation}
 {\cal A}_{2n}=\frac{1}{\pi}\int_{-\pi}^{\pi}du\,Q_p(u,0)\ce_{2n} \left( \frac{u}{2},\frac{2p}{D_R} \right). 
 \end{equation}
Note that since we start with $\theta(0)$, chosen uniformly from $[0,2\pi]$, $Q_p(u,0)=1/(2\pi)$.
At large times, i.e., $D_R t\gg 1$, the generating function $Q_p(u,t)$ in \eref{abp:qput} is dominated by the smallest eigenvalue corresponding to $n=0$. Thus at large times we have, 
\begin{align}
Q_p(u,t)\simeq {\cal A}_{0}\, \ce_{0}\left( \frac{u}{2},\frac{2p}{D_R} \right) \exp\left[-\frac{t D_R}{4}\, a_{0}\left(\frac{2p}{D_R}\right)\right].
\label{abp:qput_LT}
\end{align} 

 Thus, the generating function of the position distribution $K(p,t)$ at large times can be obtained by integrating \eref{abp:qput_LT} over $u$ as,
 \begin{align}
 K(p,t)&=\int_{-\pi}^{\pi}du~Q_p(u,t)\cr
 &=\frac{1}{2\pi^2}\left[\int_{-\pi}^{\pi}du \, \ce_{0} \left( \frac{u}{2},\frac{2p}{D_R} \right)  \right]^2 \exp\left[-\frac{t D_R}{4}\, a_{0}\left(\frac{2p}{D_R}\right)\right].\label{kpt}
 \end{align}
 The large time marginal position distribution can then be easily obtained as,
 \begin{align}
 P(w,t)=\frac{1}{2\pi i}\int_{-i\infty}^{i\infty}dp\, K(p,t) e^{pw}=\frac{1}{2\pi}\int_{-\infty}^{\infty}d\lambda
 \, K(i\lambda,t) e^{i\lambda w}.
 \end{align}
In terms of the scaled position $z=x/\sqrt{2D_\ab t}$ (used in the analysis in main text) we have,
\begin{align}
P(z,t)=\frac{1}{2\pi}\int_{-\infty}^{\infty}d\phi\, K\left(i\phi \sqrt{\frac{D_R}{2t}},t\right)\, e^{i\phi z}.\label{kpt_pzt}
\end{align}
 where 
 $\phi=\lambda \sqrt{2t/D_R}$. Thus, we need to evaluate $K(p,t)$ in \eref{kpt} and use it in \eref{kpt_pzt} to obtain the large time behavior of $P(z,t)$. The Mathieu functions $\ce_{2n}(z,q)$ can be expressed in Fourier series as,
 \begin{align}
 \ce_{2n}(z,q)=\sum_{m=0}^{\infty}E_{2m}^{2n}(q)\cos(2mz).\label{ce2n}
 \end{align}
 Using this form in \eref{kpt} simplies
 $K(p,t)$ to,
 \begin{align}
 K(p,t)=\frac{1}{2\pi^2} [2\pi E^0_0(u/2)]^2 \exp\left[-\frac{t D_R}{4}\, a_{0}\left(\frac{2p}{D_R}\right)\right].
 \label{kpt_final}
 \end{align}
Now, we need to calculate the eigenvalue $a_0$ and the Fourier coefficient of the Mathieu function i.e., $A^0_0(q)$. To do so, we note that,
\begin{align}
E^0_{2s}(q)=\left(\alpha(s)q^s+O(q^{s+2})\right)E^0_0(q),\label{Mathieu0}
\end{align}
and 
\begin{subequations}
\label{Mathieu__}
\begin{align}
a_0 E_0^0-q E_2^0&=0,\label{Mathieua}\\
(a_0-4)E_2^0-q(E_4^0+2E_0^0)&=0,\label{Mathieub}\\
(a_0-4m^2)-q(E_{2m+2}^0+E_{2m-2}^0)&=0~~m\geq 2.\label{Mathieuc}  
\end{align} 
\end{subequations}
We will first evaluate the eigenvalue $a_0$ order by order by considering it to be of the form,
\begin{equation}
a_0(q)=\sum_{n=0}^{\infty}a^{2n}_0 q^{2n}.\label{Meigenvalues}
\end{equation}
Putting this in \eref{Mathieua}, 
\begin{equation}
E_2^0(q)=\frac{a^0_0 E_0^0}{q}+O(q).
\end{equation}
 Now, to satisfy \eref{Mathieu0}, the coefficient of $q^{-1}$, $a^0_0=0$. Now, using the updated series for $a_0$ in \eref{Mathieub}, we get,
 \begin{align}
 E_4^0(q)=-2(1+2a^2_0)E_0^0+O(q^2).
 \end{align}   
Again, to satisfy \eref{Mathieu0} $a^2_0=-1/2$. Using the updated series for $a_0$ in \eref{Mathieuc} (for $m=2$), we get,
\begin{align}
E^6_0(q)=-q\left(\frac{7}{2}+64a^4_0\right)E_0^0+O(q^3),
\end{align}
leading to $a^4_0=7/128$. Proceeding similarly we can systematically evaluate $a_0$ order by order as a power series in $q$. For our purpose it is enough to use,
\begin{align}
a_0(q)=-\frac{1}{2}q^2+\frac{7 }{128}q^4-\frac{29 }{2304}q^6+\frac{68687}{18874368}q^8-\frac{123707 }{104857600}q^{10}+\cdots\label{a0_mathieu}
\end{align}

Now we use this form of $a_0$ in \eref{Mathieu__} to evaluate $E_{2m}^0$s in terms of $E_0^0$. We again state an identity, obtained by squaring both sides of \eref{ce2n} for $n=0$,
\begin{align}
2E_0^0(q)^2+\sum_{m=1}^{\infty}\Big[E_{2m}^0(q)\Big]^2=1.
\label{mathieu_norm}
\end{align} 

 Again, from \eref{Mathieu__} we can find $E_{2m}^0(q)$s in terms of $E_0^0(q)$, for example,
\begin{subequations}
\begin{align}
E_2^0(q)&=\frac{a_0}{q} E_0^0,\\
E_4^0(q)&=\left[ \frac{a_0(a_0-4)}{q^2}-2\right] E_0^0,\\
E_6^0(q)&=\left[ \frac{a_0(a_0-4)(a_0-16)}{q^3}-\frac{3a_0-32}{q}\right] E_0^0,\\
E_8^0(q)&=-\left[\frac{a_0(a_0-4)(a_0-16)(a_0-36)}{q^4}+\frac{4(a_0-12)(a_0-24)}{q^2}-2\right] E_0^0,\\
E_{10}^0(q)&=\left[\frac{a_0(a_0-4)(a_0-16)(a_0-36)(a_0-64)}{q^5}+\frac{(73728 - a_0 (10432 + 5a_0 (a_0-84))) }{q^3}\right.\cr
&\qquad\qquad\left.+\frac{5(a_0-32)}{q}\right]E_0^0.\nonumber
\end{align}
\end{subequations}
We can use the above expressions in \eref{mathieu_norm} to obtain $E_0^0(q)$. Finally, we arrive at,
\begin{align}
E_0^0(q)=&\frac{1}{\sqrt{2}}-\frac{q^2}{16 \sqrt{2}}+\frac{79 q^4}{4096 \sqrt{2}}-\frac{36919 q^6}{5308416 \sqrt{2}}+O(q^8).\label{A00_mathieu}
\end{align}

Note that, to obtain $E_0^0$ to $O(q^{2m})$, we need to keep upto $E_{4m}^0(q)$, for which we need to compute $a_0(q)$ upto $O(q^{4m})$.
Now, using $a_0(q)$ and $E_0^0(q)$ (from \eref{a0_mathieu} and \eref{A00_mathieu} respectively) in \eref{kpt_final} and expanding as a power series in $t$ as $t\to\infty$, we get,
\begin{align}
K(p,t)=&e^{-\frac{p^2}{4}} \left(1+  \frac{ p^2 (32-7 p^2)}{128 D_R t}+\frac{p^4 (4608 - 11456 p^2 + 441 p^4)}{294912 D_R^2 t^2}\right.\cr
&\left.+\frac{p^6 (44357632 - 10240632 p^2 + 594720 p^4 - 9261 p^6)}{339738624 D_R^3t^3}+O(t^{-4}) \right).
\end{align}
The above expression upon Fourier transformation yields,
\begin{align}
P(z,t)&=\frac{e^{-z^2}}{\sqrt{\pi }}\left[1-\frac{1}{D_R t}\left(\frac{5}{32}-\frac{13z^2}{8}+\frac{7 z^4}{8}\right)\right.\cr 
&\left.+ \frac{1}{(D_R t)^2}\left(
\frac{49 z^8}{128}-\frac{1655 z^6}{576}+\frac{3203 z^4}{768}-\frac{73 z^2}{256}-\frac{677}{6144}\right)\right.\cr
&\left.
+\frac{1}{(D_Rt)^3}\Big(-\frac{343 z^{12}}{3072}+\frac{17437 z^{10}}{9216}-\frac{325727 z^8}{36864}+\frac{1790095 z^6}{165888}\right.\cr
&\left.+\frac{293635 z^4}{442368}-\frac{92375 z^2}{147456}-\frac{302975}{1769472}\Big)
 +O(t^{-4})\right].
\end{align}
 The obtained terms agree with the corrections obtained in the main text \eref{abp:q2F}, \eref{abp:q4F} and \eref{abp:q6final}  using our perturbative strategy.

 \section{Intermediate steps in the computation of $A_0^4(y,t)$ and $A_0^6(y,t)$ for DRABP}\label{app:drabpdetails}
In this section, we provide the intermediate steps leading to the subleading contributions $A_0^4(y,t)$ and $A_0^6(y,t)$ for DRABP.  
 
Setting $k=6$ in \eref{drabp:a0k}, we get an inhomogeneous diffusion equation for $A_0^4(y,t)$,
\begin{align}
\Bigg[\frac{\partial }{\partial t}-\frac{\partial^2}{\partial y^2}\Bigg]A_0^4(y,t)=S_{4}(y,t),
\end{align}
where the inhomogeneous part is given by,
\begin{align}
S_4(y,t)=\frac{\partial^2}{\partial  y^2}&\Bigg[\left(-\frac{1}{\lambda+1}\frac{\partial}{\partial t}+\frac{1}{8}\frac{\partial ^2}{\partial y^2}
\right)A_0^2+\left(\frac{1}{{(\lambda +1)^2}}\frac{\partial ^2}{ \partial t^2}-\frac{(\lambda+9)}{32(\lambda +1)}\frac{\partial ^3}{\partial y^2\, \partial t}\right.\cr
&\left.+\frac{(\lambda+5)}{32(\lambda+9)}\frac{\partial^4}{\partial y^4}\right)A_0^0 \Bigg].
\end{align}
Considering the scaling form for $A_0^4(y,t)$, as given by \eref{drabp:q2kgen} and using the explicit forms of $A_0^0(y,t)$ and $A_0^2(y,t)$, we have get an inhomogeneous Hermite equation for $q_4(z)$ like in~\eqref{gen:hermite_inh}. The inhomogeneous term is given by,
%\begin{align}
%A_0^4(y,t)=\frac{e^{-y^2/(4t)}}{\sqrt{4\pi t}}\frac{1}{t^2}q_4\left(\frac{y}{\sqrt{4t}}\right),
%\end{align}
%where $q_4(z)$ satisfies,
%\begin{align}
%q_4''(z)-2zq_4'(z)+8q_4(z)=s_4(z),
%\label{de:q4:drabp}
%\end{align}
\begin{align}
s_4(z)=\sum_{n=0}^{4}r_{4,2n}(\lambda)z^{2n},
\end{align}
where the coefficients $\{r_{4,n}\}$ are,
\begin{subequations}
\begin{align}
r_{4,0}(\lambda)&=-\frac{15}{256(\lambda +1)^2 (\lambda +9)}(223-251 \lambda+45 \lambda ^2+7 \lambda ^3 ),\cr
r_{4,2}(\lambda)&=\frac{15}{32(\lambda +1)^2(\lambda +9) }(939-323 \lambda+25 \lambda ^2 +7 \lambda ^3),\cr
r_{4,4}(\lambda)&=-\frac{15}{32(\lambda +1)^2(\lambda +9) }(1655-395 \lambda+5 \lambda ^2+7 \lambda ^3 ),\cr
r_{4,6}(\lambda)&=\frac{1}{8(\lambda +1)^2(\lambda +9) }(2371-467 \lambda-15 \lambda ^2+7 \lambda ^3),\cr
r_{4,8}(\lambda)&=-\frac{1}{16(\lambda +1)^2(\lambda +9) }(441-77 \lambda-5 \lambda ^2+\lambda ^3 ).\nonumber
\end{align}
\end{subequations}
The general solution for $q_4(z)$ can be again obtained using \eref{eq:qn_gen} in terms of an undetermined constant $C_4$. This constant can be found out by comparing the coefficient of $(\tau/t)^2$ in the expansion of $M(4,0,t)/(4t)^2$ of \eref{drabp:m4t} to the one obtained from the approximate distribution \eref{drabp:marginal} (more precisely \eqref{drabp:m2k_series} with $k=2$). Following this procedure we finally get,
\begin{align}
q_4(z)=\sum_{n=0}^{4}\alpha_{4,n}(\lambda)z^{2n},
\label{app:drabpq4_1}
\end{align}
with
\begin{align}
\alpha_{4,0}&=-\frac{1}{2048 (\lambda +1)^2 (\lambda +9)}(2031-411 \lambda-963 \lambda ^2-57 \lambda ^3),\cr
\alpha_{4,1}&=-\frac{1}{256 (\lambda +1)^2 (\lambda +9)}(219-559\lambda+273\lambda^2+27\lambda^3),\cr
\alpha_{4,2}&=\frac{1}{256 (\lambda +1)^2 (\lambda +9)}(9609-3789\lambda+523\lambda^2+97\lambda^3),\cr
\alpha_{4,3}&=\frac{1}{64 (\lambda +1)^2 (\lambda +9)}(1655-395\lambda+5\lambda^2+7\lambda^3),\cr
\alpha_{4,4}&=\frac{1}{128 (\lambda +1)^2 (\lambda +9)}(441-77\lambda-5\lambda^2+\lambda^3).
\label{app:drabpq4_2_}
\end{align}

Similarly, we can find an inhomogeneous diffusion equation for the next subleading order contribution $A_0^6(y,t)$, by setting $k=8$ in \eref{drabp:a0k},
\begin{align}
\Bigg[\frac{\partial }{\partial t}-\frac{\partial^2}{\partial y^2}\Bigg]A_0^6(y,t)=S_{6}(y,t),
\end{align}
where the inhomogeneous term is given by,
\begin{align}
\fl S_{6}(y,t)&=\frac{\partial ^2}{\partial y^2}\Bigg[\Big(-\frac{1}{\lambda+1} \frac{\partial}{\partial t}+\frac 18\frac{\partial ^2}{\partial y^2}\Big)A_0^4\cr
&+\Big(\frac{1}{(\lambda+1)^2}\frac{\partial ^2}{\partial t^2}-\frac{\lambda+9}{32(\lambda+1)}\frac{\partial ^3}{\partial t\partial y^2}+\frac{\lambda+5}{32(\lambda+9)}\frac{\partial^4 }{\partial y^4}   \Big)A_0^2\cr
&-\Bigg(\frac{1}{(\lambda+1)^3}\frac{\partial^3}{\partial t^3}-\frac{57+10\lambda+\lambda^2}{128(\lambda+1)^2} \frac{\partial^4}{\partial y^2\partial t^2}	+\frac{307+135\lambda+21\lambda^2+\lambda^3}{64 (\lambda +1) (\lambda +9)^2}\frac{\partial^5}{\partial y^4\partial t}\cr
&-\frac{481+162\lambda+17\lambda^2}{2048(\lambda+9)^2}\frac{\partial^6}{\partial y^6}\Bigg)A_0^0\Bigg].
\end{align}
Again, considering the scaling form for $A_0^6(y,t)$, as given by \eref{drabp:q2kgen} and using the explicit forms of $A_0^0(y,t)$, $A_0^2(y,t)$ and $A_0^4(y,t)$, we have get an inhomogeneous Hermite equation for $q_6(z)$ like in~\eqref{gen:hermite_inh}. The inhomogeneous term is given by,
%Considering $A_0^6(y,t)$ to be of the form,
%\begin{align}
%A_0^6(y,t)=\frac{e^{-y^2/(4t)}}{\sqrt{4\pi t}}\frac{1}{t^3}q_6\left(\frac{y}{\sqrt{4t}}\right),
%\end{align}
%where $q_6(z)$ satisfies,
%\begin{align}
%q_6''(z)-2zq_6'(z)+8q_6(z)=s_6(z),
%\label{de:q6:drabp}
%\end{align}
\begin{align}
s_6(z)=\sum_{n=0}^{6}r_{6,2n}(\lambda)z^{2n}.
\end{align}
The coefficients $\{r_{6,n}\}$ are given by,
%\begin{align}
\bea
\fl r_{6,0}(\lambda)=\frac{ -4389525+2500575 \lambda+2973390 \lambda ^2-626850 \lambda ^3-158025 \lambda ^4-7245 \lambda ^5}{16384 (\lambda +1)^3 (\lambda +9)^2},\cr
\fl r_{6,2}(\lambda)=\frac{979965 + 13285545 \lambda - 11694270 \lambda^2 + 480690 \lambda^3 + 420945 \lambda^4 + 
 24885 \lambda^5}{4096 (\lambda +1)^3 (\lambda +9)^2},\cr
\fl r_{6,4}(\lambda)=\frac{108286605 - 63628215 \lambda - 7444304 \lambda^2 - 1734096 \lambda^3 - 585375 \lambda^4 - 
 45675 \lambda^5}{4096 (\lambda +1)^3 (\lambda +9)^2},\cr
\fl r_{6,6}(\lambda)=\frac{-20520801+7463603 \lambda-930538 \lambda^2-216762 \lambda^3+29435 \lambda^4+3255 \lambda^5}{512 (\lambda +1)^3 (\lambda +9)^2},\cr
\fl r_{6,8}(\lambda)=\frac{17055513 - 4774859 \lambda + 108634 \lambda^2 + 142026 \lambda^3 - 7715 \lambda^4 - 1455 \lambda^5}{1024 (\lambda +1)^3 (\lambda +9)^2},\cr
\fl r_{6,10}(\lambda)=\frac{-619479 + 145477 \lambda + 7162 \lambda^2 - 4118 \lambda^3 + 45 \lambda^4 + 33 \lambda^5}{256 (\lambda +1)^3 (\lambda +9)^2},\cr
\fl r_{6,12}(\lambda)=\frac{27783 - 5733\lambda  - 602 \lambda^2 + 150 \lambda^3 + 3 \lambda^4 - \lambda^5}{256 (\lambda +1)^3 (\lambda +9)^2}.
\eea
%\end{align}
The general solution for $q_6(z)$ can be again obtained using \eref{eq:qn_gen} in terms of an undetermined constant $C_6$. This constant can be found out by comparing the coefficient of $(\tau/t)^2$ in the expansion of $M(6,0,t)/(4t)^3$ of \eref{drabp:m4t} to the one obtained from the approximate distribution \eref{drabp:marginal} (more precisely \eqref{drabp:m2k_series} with $k=3$). Following this procedure we finally get,
\begin{align}
q_6(z)=\sum_{n=0}^{6}\alpha_{6,n}(\lambda)z^{2n},
\label{app:drabpq6_1}
\end{align}
where the coefficients $\{\alpha_{6,n}(\lambda)\}$,
\begin{align}
\alpha_{6,0}(\lambda)&=\frac{-908925 - 261945\lambda+ 444030 \lambda^2 + 
 311790 \lambda^3 + 23535 \lambda^4 + 315 \lambda^5}{65536(\lambda +1)^3 (\lambda +9)^2},\cr
\alpha_{6,1}(\lambda)&= \frac{-831375 + 1643205 \lambda + 820650 \lambda^2 -  781110 \lambda^3 - 114315 \lambda^4 - 4095 \lambda^5}{16384(\lambda +1)^3 (\lambda +9)^2},\cr
\alpha_{6,2}(\lambda)&=\frac{880905 + 3333045 \lambda - 4445190 \lambda^2 +  680970 \lambda^3 + 216525 \lambda^4 + 11025 \lambda^5}{16384(\lambda +1)^3 (\lambda +9)^2},\cr
 \alpha_{6,3}(\lambda)&=\frac{1790095 - 1116021 \lambda + 393878 \lambda^2 +  10774 \lambda^3 - 13365 \lambda^4 - 945 \lambda^5}{2048(\lambda +1)^3 (\lambda +9)^2},\cr
 \alpha_{6,4}(\lambda)&=\frac{-2931543 + 1066229 \lambda - 132934 \lambda^2 -  30966 \lambda^3 + 4205 \lambda^4 + 465 \lambda^5}{4096(\lambda +1)^3 (\lambda +9)^2},\cr
 \alpha_{6,5}(\lambda)&=\frac{156933 - 41207 \lambda - 270 \lambda^2 + 1234 \lambda^3 -  39 \lambda^4 - 11 \lambda^5}{1024(\lambda +1)^3 (\lambda +9)^2},\cr
 \alpha_{6,6}(\lambda)&=\frac{-27783 + 5733 \lambda + 602 \lambda^2 - 150 \lambda^3 -  3 \lambda^4 + \lambda^5}{3072 (\lambda +1)^3 (\lambda +9)^2}.
 \label{app:drabpq6_2}
\end{align}
The subleading order contributions obtained above are compared with numerical simulations in \ref{F:drabp} and show good agreement.

\section*{References}
\bibliography{ref}

\end{document}